\documentclass[superscriptaddress,aps,prd,showpacs,preprint]{revtex4-1}

\usepackage{eurosym}
\usepackage{graphicx}
\usepackage{amsmath}
\usepackage{mathrsfs}
\usepackage{bm}
\usepackage{color}
\usepackage{amsmath}
\usepackage{amssymb}
\usepackage[section]{placeins}

\usepackage{multirow}
\usepackage{array}
\usepackage[caption=false]{subfig}
\def\be{\begin{equation}}
\def\ee{\end{equation}}
\def\bea{\begin{eqnarray}}
\def\eea{\end{eqnarray}}

\allowdisplaybreaks

\begin{document}

\title{Massive Scalar Perturbations on Myers-Perry-de Sitter Black Holes with a Single Rotation }

\author{Supakchai Ponglertsakul}
\email{supakchai.p@gmail.com}
\affiliation{Division of Physics and Semiconductor Science, Dongguk University, Seoul 04620,
Republic of Korea}
\author{Bogeun Gwak}
\email{rasenis@dgu.ac.kr}
\affiliation{Division of Physics and Semiconductor Science, Dongguk University, Seoul 04620,
Republic of Korea}

\newpage

\begin{abstract}
This study investigates the stability of higher-dimensional singly rotating Myers-Perry-de Sitter (MP-dS) black holes against scalar field perturbations. The phase spaces of MP-dS black holes with one spin parameter are discussed. Additionally, the quasinormal modes (QNMs) of MP-dS black holes are calculated via the asymptotic iteration method and sixth-order Wentzel--Kramers--Brillouin approximation. For near-extremal MP-dS black holes, the event horizon may be considerably close to the cosmological horizon. In such cases, the P\"oschl-Teller technique yields an accurate analytic formula for the QNMs. It is found that when the spin parameter of a black hole increases, the scalar perturbation modes oscillate at higher frequencies and decay faster. Furthermore, the MP-dS black hole with a single rotation is found to be stable under perturbation.
\end{abstract}

\maketitle

\section{Introduction}\label{sec1}

Black holes are compact objects that describe the final state of massive stellar bodies in our universe. They are represented by the exact solutions of Einstein's field equations of general relativity. These solutions provide simple but useful tools to probe physics in a strong gravity regime. The most well-known exact solution, i.e., the Schwarzschild metric, represents an exterior geometry of static, spherically symmetric matter. A generalization of the Schwarzschild metric with a rotating effect yields the Kerr metric \cite{Kerr:1963ud}. The Kerr solution describes astronomical rotating bodies, including black holes and neutron stars. Recently, the detection of gravitational waves \cite{TheLIGOScientific:2016src} and images of black hole shadows \cite{Akiyama:2019cqa} have confirmed the effectiveness of the Kerr solution in describing astronomical black holes. 

Owing to string theory and brane-world models, higher-dimensional gravity has attracted interest as a possible framework for the unified theory. Notably, the exact solutions of Einstein's field equation can be generalized to higher dimensions. In fact, string theory and brane-world scenarios often predict the existence of higher-dimensional black holes. In the four-dimensional case, uniqueness theorem restricts any stationary black holes in an asymptotically flat spacetime to being described by the Kerr solution. However, a similar statement for higher-dimensional cases has not been found, and the development of a black hole uniqueness theorem in $d$ dimensions remains a work in progress \cite{Hollands:2012xy}. In addition, several higher-dimensional black objects have a variety of non-trivial topologies. Therefore, various solutions exist for $d>4$, such as black branes, strings, and rings \cite{Emparan:2008eg}. Another area wherein higher-dimensional black holes have proven useful is anti-de Sitter/conformal field theory (AdS/CFT) correspondence \cite{Maldacena:1997re}. The $d$ dimensional black holes with a negative cosmological constant have emerged as a powerful tool for probing non-perturbative regimes of strongly coupled $(d-1)$ dimensional quantum field theories. Thus, since the advent of AdS/CFT correspondence, exact solutions for Einstein's gravity in higher dimensions with a cosmological constant have attracted significant attention.

The classical stability of a given solution is an important property, and therefore should be examined first. As an astronomical object, a black hole in nature is likely to be found in a stable state. In four dimensions, black holes are generally stable under gravitational perturbations. However, the stability behaviors of multidimensional black holes are significantly complex. Black strings and p-branes generally suffer from Gregory-Laflamme instability \cite{Gregory:1993vy}, whereas Tangherlini metric \cite{Ishibashi:2003ap} and $d$ dimensional Reissner-Nordstr\"om (RN) \cite{Kodama:2003ck} black holes are gravitationally stable. With non-vanishing cosmological constants, multidimensional black holes are shown to be stable \cite{Konoplya:2007jv}, including the RN-AdS under gravitational perturbations \cite{Konoplya:2008rq}.  However, the RN-dS is unstable for $d\geq 7$ \cite{Konoplya:2008au}. With the presence of superradiance effect (for a nice review on this subject sees \cite{Brito:2015oca}), charged scalar field propagating on higher dimensional RN-dS suffers from superradiant instability \cite{Destounis:2019hca}.

Compared to static cases, the stability of $d>4$ rotating black holes has not been investigated sufficiently. This is because decoupling the perturbation equations is significantly more complicated in the latter case. The most well-known study of classical stability of higher-dimensional rotating bodies focuses on the Myers-Perry (MP) black hole. This is a black hole solution with arbitrary rotation in $N=\frac{d-1}{2}$ possible independent rotation planes \cite{Myers:1986un}. It has been shown that the perturbation equation describing a massless scalar field on the MP black hole with a single rotation axis can be separated into arbitrary dimensions \cite{Ida:2002zk}. The singly rotating black holes in $d=5$ and $d=6$ are proven to be stable against scalar perturbations \cite{Ida:2002zk,Cardoso:2004cj}, even in the ultra-spinning regime \cite{Morisawa:2004fs}. For $d\geq 7$, the simply rotating MP black holes are stable against tensorial perturbations \cite{Kodama:2009bf}. In contrast, MP black holes in asymptotic AdS generally suffer from superradiant instability. MP-AdS black holes in odd dimensions ($d>5$) with equal angular momentum are superradiantly unstable against gravitational perturbations \cite{Kunduri:2006qa}. With two independent rotation parameters, five-dimensional charged rotating black holes in AdS spacetime are only unstable to modes having an even orbital quantum number \cite{Aliev:2008yk}. In \cite{Kodama:2009rq}, gravitational perturbations of rapidly rotating MP-AdS black holes with a single rotating parameter are considered and shown to be unstable. Furthermore, superradiant instability has been analytically proven for small singly rotating MP-AdS black holes in arbitrary dimensions \cite{Delice:2015zga}. 

For a positive cosmological constant, the Kerr-dS black holes in general relativity are gravitationally stable \cite{Yoshida:2010zzb}, while their counterparts in scalar tensor theory suffer from superradiant instability under scalar perturbations \cite{Zhang:2014kna}. Recently, the quasinormal modes (QNMs) of higher-dimensional singly rotating Myers-Perry-de Sitter (MP-dS) black holes with non-minimally coupled scalar fields were analytically studied, and a near-extremal formula for quasinormal frequencies was derived \cite{Gwak:2019ttv}. 
The proposed study extends the work of \cite{Gwak:2019ttv} by considering massive scalar perturbations on singly rotating MP-dS black holes. In Sect~\ref{sec:setup}, the MP-dS metric is introduced and behavior of their phase spaces is discussed. In Sect~\ref{sec:KGeq}, the Klein-Gordon equation is separated into radial and angular equations. The superradiant condition and quasinormal boundary conditions are given; further, the series expansion of the separation of the constant is derived within a slowly rotating limit. In Sect~\ref{sec:nearextremal}, the near-extremal condition for the quasinormal frequencies are derived. The calculation methods used herein for the sixth-order Wentzel--Kramers--Brillouin (WKB) approximation and asymptotic iteration method (AIM) are given in Sect~\ref{sec:wkb} and Sect~\ref{sec:AIM}, respectively. Sect~\ref{sec:results} presents the calculation and discussion of the QNMs. Finally, the results are summarized in Sect~\ref{sec:conclu}.

\section{Myers-Perry-de Sitter Black Holes with a Single Rotation}\label{sec:setup}

\begin{figure}[h]
\centering
\includegraphics[width=0.4\textwidth]{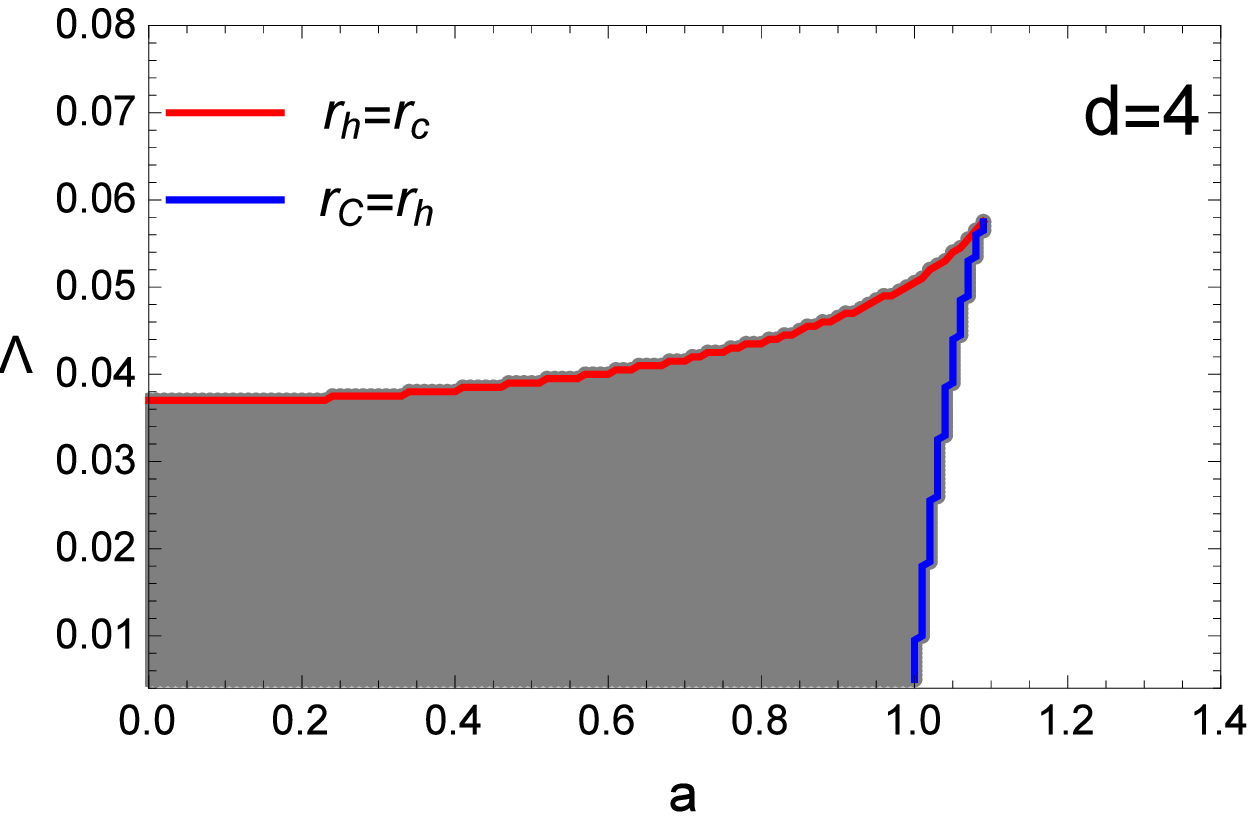}\quad 
\includegraphics[width=0.4\textwidth]{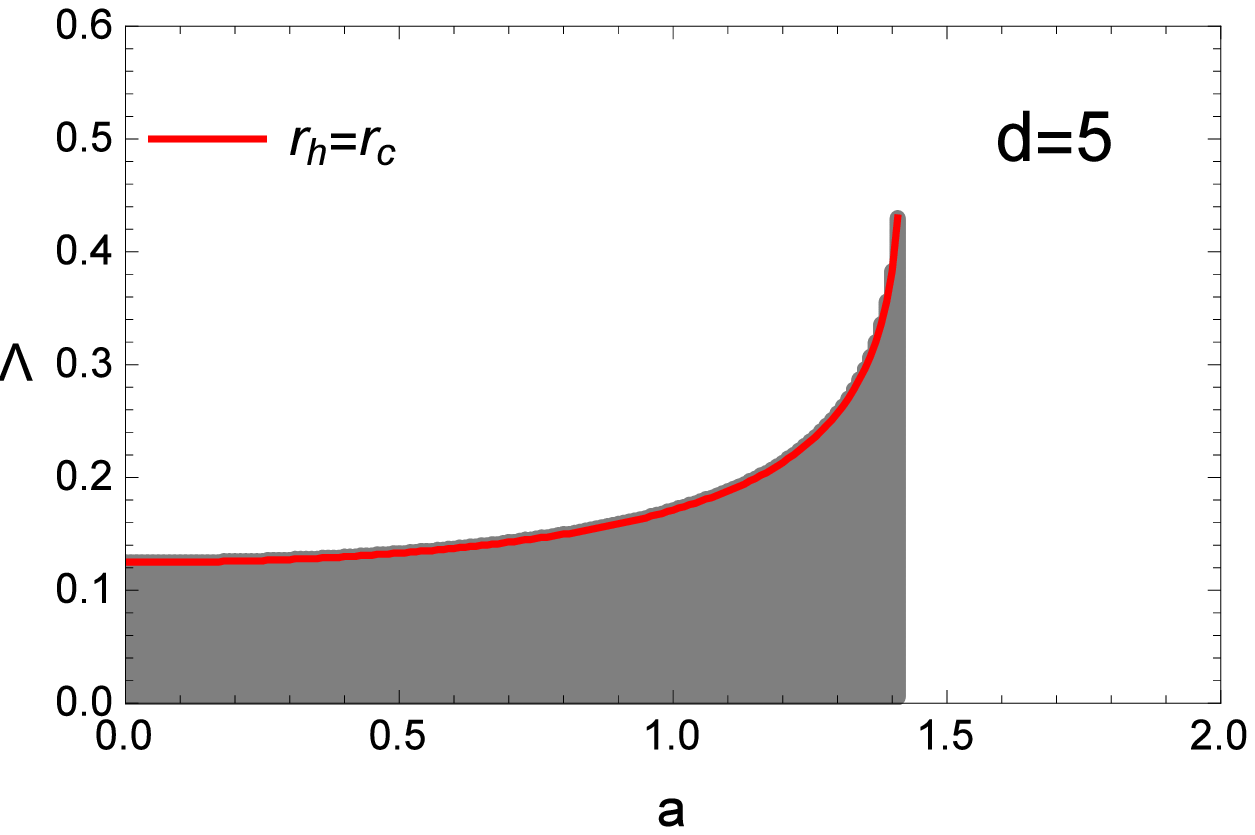}\\
\includegraphics[width=0.4\textwidth]{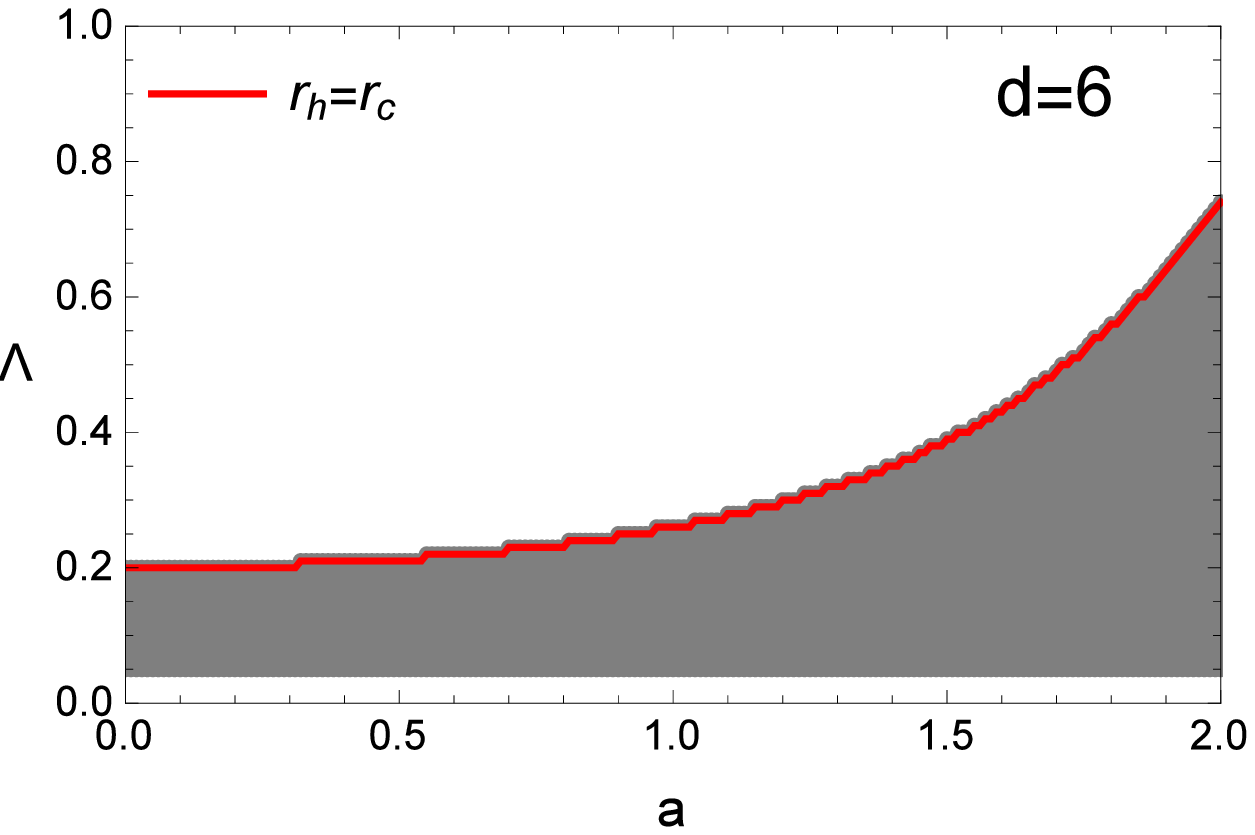}\quad
\includegraphics[width=0.4\textwidth]{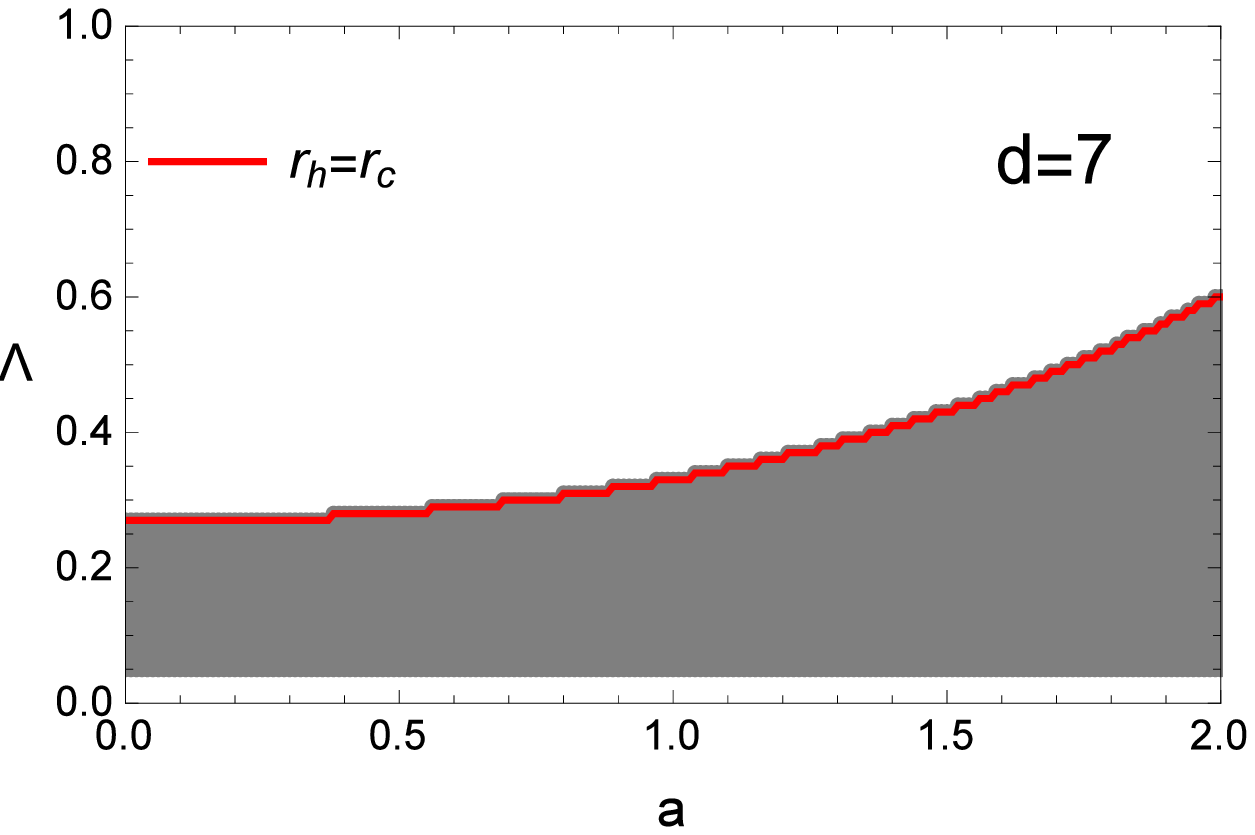}
\caption{Phase spaces of singly rotating black holes in de Sitter space in various dimensions with $M=1$.} \label{phasespace}
\end{figure}

The MP black hole is the generalized Kerr black hole including multiple angular momenta for d-dimensional Einstein's gravity \cite{Myers:1986un}. It is extended to a version that includes the positive cosmological constant \cite{Gibbons:2004uw,Gibbons:2004js}. Through this positive cosmological constant, the asymptotic geometry of the black hole represents the dS spacetime. The proposed study investigates the MP-dS black hole with a single rotation. Hence, the black hole has two conserved quantities, i.e., mass and angular momentum. The metric with the cosmological constant $\Lambda$ is given as
\begin{align}\label{metric}
ds^2 &= -\frac{\Delta_r}{\rho^2} \left(dt - \frac{a \sin^2\theta}{\Sigma}d\phi\right)^2 + \frac{\rho^2}{\Delta_r}dr^2 + \frac{\rho^2}{\Delta_{\theta}}d\theta^2 + \frac{\Delta_{\theta}\sin^2\theta}{\rho^2}\left(a dt - \frac{\left(r^2+a^2\right)}{\Sigma}d\phi\right)^2 \\ \nonumber 
&~~~+ r^2\cos^2\theta d\Omega^2_{d-4},
\end{align}
where
\begin{align}
\rho^2 &= r^2 + a^2\cos^2\theta,~~~~\Delta_r = \left(r^2+a^2\right)\left(1-\Lambda r^2\right)-2Mr^{5-d}, \\ \nonumber 
\Sigma &= 1 + \Lambda a^2,~~~~~~~~~~~\Delta_{\theta} = 1 + \Lambda a^2\cos^2\theta.
\end{align}
Only one angular momentum is considered; thus, the other rotational planes remain empty spatial $(d-4)$ dimensions. They are represented by the $(d-4)$ sphere in the metric.
\begin{align}
d\Omega^2_{d-4} &= \sum_{i=1}^{d-4}\left(\prod_{j=1}^{i}\sin^2\psi_{j-1}\right)d\psi^2_{i},~~~~\psi_0 \equiv \frac{\pi}{2}.
\end{align}
Three horizons are located at the points satisfying $\Delta_r=0$. These are the inner, outer, and cosmological horizons, denoted by $r_\text{C}$, $r_\text{h}$, and $r_\text{c}$, respectively. Mass and spin parameters are denoted by $M$ and $a$, which are associated with mass and angular momentum of the black hole, respectively \cite{Altamirano:2013ane}.
\begin{align}
M_\text{B}=\frac{\Omega_{d-2}}{4\pi}\frac{M}{\Sigma^2}\left(1+\frac{(d-4)\Sigma}{2}\right),\quad J_\text{B}=\frac{\Omega_{d-2}}{4\pi}\frac{Ma}{\Sigma^2}.
\end{align}
However, the metric (\ref{metric}) continues to rotate at the limit of $r\rightarrow \infty$. This makes it challenging to define conserved scalar field quantities, such as the energy and angular number. Hence, the conserved quantities can be determined at the static boundary, which is produced by
\begin{align}\label{eq:transformations1}
dt \rightarrow dT, \quad d\phi \to d\Phi + a\Lambda dT.
\end{align}
The transformed metric is within the static asymptotic boundary \cite{Hawking:1998kw}. Then, the transformed metric is 
\begin{align}
ds^2 &= -\frac{\Delta_r}{\rho^2\Sigma^2} \left(\Delta_{\theta}dT - a \sin^2\theta d\Phi\right)^2 + \frac{\rho^2}{\Delta_r}dr^2 + \frac{\rho^2}{\Delta_{\theta}}d\theta^2 + r^2\cos^2\theta d\Omega^2_{d-4} \nonumber \\ 
&\quad + \frac{\Delta_{\theta}\sin^2\theta}{\rho^2\Sigma^2}\left(a\left(1-\Lambda r^2\right) dT - \left(r^2+a^2\right)d\Phi\right)^2. \label{metric1}
\end{align}
Coordinates transformations (\ref{eq:transformations1}) do not change $\Delta_r$ (\ref{metric}), therefore, the three horizons remain unaffected. The angular velocities at each corresponding horizons are modified to
\begin{align}
\Omega_\text{C} = \frac{a\left(1-r_\text{C}^2\Lambda\right)}{r_\text{C}^2+a^2},\quad \Omega_\text{h} = \frac{a\left(1-r_\text{h}^2\Lambda\right)}{r_\text{h}^2+a^2}, \quad \Omega_\text{c} = \frac{a\left(1-r_\text{c}^2\Lambda\right)}{r_\text{c}^2+a^2}.
\end{align}
Moreover, the surface gravities at the horizons are given by
\begin{align}\label{surfacegravity}
\kappa_\text{C} = \frac{1}{2\left(r^2_\text{C}+a^2\right)}\frac{d\Delta_{r}(r)}{dr}{\Big|}_{r=r_\text{C}},\quad \kappa_\text{h} = \frac{1}{2\left(r^2_\text{h}+a^2\right)}\frac{d\Delta_{r}(r)}{dr}{\Big|}_{r=r_\text{h}},\quad \kappa_\text{c} = \frac{1}{2\left(r^2_\text{c}+a^2\right)}\frac{d\Delta_{r}(r)}{dr}{\Big|}_{r=r_\text{c}}.
\end{align}
Because we considers the spacetime with three horizons in higher dimensions, two extremal conditions are set: $r_\text{C}=r_\text{h}$ and $r_\text{h}=r_\text{c}$. Furthermore, the number of horizons and possible extremal conditions depends on the dimensionality and number of rotational planes considered. Figure \,\ref{phasespace} shows the phase spaces of the black holes in the case of single rotation. For a given mass, the black hole phase spaces are determined by the cosmological constant and spin parameter.
In four dimensions, two types of extremal black holes are possible. The phase space of black holes are confined by these two extremal conditions. In five dimensions, only the condition $r_\text{h}=r_\text{c}$ is possible. The existence of black hole with three horizons is prohibited by the certain value of spin parameter. However, beyond five dimensions, there are no limits in the spin parameter, and $r_\text{h}=r_\text{c}$ extremal condition still applies. Therefore, by scalar perturbation, the stabilities of the various black holes can be investigated with respect to dimensionality, cosmological constant, and spin parameter.

\section{Massive Scalar Field Equation}\label{sec:KGeq}

In this work, the stability of MP-dS black hole under scalar perturbation is investigated. The scalar field with mass $\mu$ propagating on curved background is governed by 
\begin{align}
\frac{1}{\sqrt{-g}}\partial_{\mu}\left(\sqrt{-g}g^{\mu\nu}\partial_{\nu}\Psi\right)-\mu^2\Psi &=0,  \label{KG}
\end{align}
where the metric determinant is given by
\begin{align}
\sqrt{-g} &=\frac{\rho^2}{\Sigma}\sin\theta \left(r\cos\theta\right)^{d-4}\prod_{j=1}^{d-5} \left(\sin\psi_j\right)^{d-4-j}.
\end{align}
The solution of the scalar field can be written as
\begin{align}
\Psi= e^{-i\omega T}e^{im\Phi}R(r)S(\theta)Y(\psi_1,\psi_2,...\psi_{d-4}).
\end{align}
The solutions to coordinates $T$ and $\Phi$ are trivial, and the separated equations with respect to $\psi_i$ are the hyperspherical harmonics about the $(d-4)$-dimensional Laplacian with an eigenvalue $-j(j+d-5)$ \cite{Boonserm:2014fja},
\begin{align}
\Delta_S Y \equiv \sum_{j=1}^{d-4} \frac{g^{\psi_j\psi_j}}{(\sin\psi_j)^{d-4-j}}\partial_{\psi_j}\left((\sin\psi_j)^{d-4-j}\partial_{\psi_j}Y\right)= -j(j+d-5)Y.
\end{align}
Then, the Klein-Gordon equation (\ref{KG}) becomes
\begin{align}
&\frac{1}{r^{d-4}}\partial_{r}\left[r^{d-4}\Delta_r \partial_r R\right] + \left[ \frac{\left(\omega(r^2+a^2)-ma\left(1-\Lambda r^2\right)\right)^2}{\Delta_r}-\mu^2 r^2 \right]R \nonumber \\
& \quad + \left[\frac{\partial_{\theta}\left[\sin\theta(\cos\theta)^{d-4}\Delta_{\theta}\partial_{\theta}S  \right]}{S\sin\theta(\cos\theta)^{d-4}}  + \left(\frac{1}{\Delta_{\theta}}\left\{2\omega m a \Delta_{\theta} -\omega^2a^2\sin^2\theta - \frac{m^2\Delta_{\theta}^2}{\sin^2\theta} \right\} -\mu^2a^2\cos^2\theta \right) \right. 
\nonumber \\
& \quad \left. -  j(j+d-5)\rho^2\right]R=0. \label{KG1}
\end{align}
By introducing separation constant $A_{kjm}$, the Klein-Gordon equation (\ref{KG1}) can be separated into the radial and angular part. The radial equation is given by
\begin{align}
 \frac{1}{r^{d-4}}\partial_r\left[r^{d-4}\Delta_r \partial_rR \right] + \left[\frac{1}{\Delta_r}\left(\omega\left(r^2 + a^2\right)-ma\left(1-\Lambda r^2\right)\right)^2 -\mu^2r^2 - \frac{j(j+d-5)a^2}{r^2}  \right. 
\nonumber \\
 \quad \left. - A_{kjm}\right]R = 0. \label{radialeq}
\end{align}
In addition, the $\theta$ equation is 
\begin{align}
\frac{1}{\sin\theta\cos^{d-4}\theta}\partial_{\theta}\left[\sin\theta\cos^{d-4}\theta\Delta_{\theta}\partial_\theta S\right] 
+ \left[-\frac{1}{\Delta_\theta}\left(a\omega\sin\theta-\frac{m\Delta_{\theta}}{\sin\theta}\right)^2 -\mu^2a^2\cos^2\theta \right. \nonumber \\
 \quad \left.  - \frac{j(j+d-5)}{\cos^2\theta} + A_{kjm}\right]S = 0. \label{angulareq}
\end{align}
The angular equation can be considered as the generalized scalar hyperspheroidal equation \cite{Cho:2009wf}. The separation constant $A_{kjm}$ is indexed by integer $k$, $j$, and $m$ \cite{Cho:2009wf,Berti:2005gp}. Furthermore, it is a function of $d,a,\omega,\Lambda$ and $\mu$.

\subsection{Radial Equation Analysis}

The radial equation becomes a Schr\"{o}dinger-like equation in terms of the tortoise coordinate $r^*$. 
\begin{align}
\frac{dr_{\ast}}{dr} = \frac{r^2+a^2}{\Delta_r},\quad R = \frac{\bar{R}}{\sqrt{r^{d-4}(r^2+a^2)}}. \label{tortoise}
\end{align}
Then, (\ref{radialeq}) can be rewritten as
\begin{align}
\frac{d^2\bar{R}}{dr_\ast^2} + \Biggl[ \left(\omega-m\Omega\right)^2 + \frac{\Delta_r}{\left(r^2+a^2\right)^2}\left(-A_{kjm}-\mu^2r^2 - \frac{j(j+d-5)a^2}{r^2} \right. \nonumber \\
 +  \left. \Delta_r \left( \frac{(2r^2-a^2)}{(r^2+a^2)^2} - \frac{(d-4)(d-6)}{4r^2} \right) - \frac{\Delta_r'}{2r} \left( \frac{2r^2}{r^2+a^2} + d-4 \right)  \right)  \Biggr]\bar{R} &= 0, \label{radialtortoise}
\end{align}
where $\Delta_r' = \frac{d\Delta_r}{dr}$. The radial range $r_\text{h}\leq r \leq r_\text{c}$ is also transformed into $-\infty < r^* < +\infty$, respectively. Then, at the radial boundaries, $r^*\rightarrow \pm \infty$, the general forms of the solutions to (\ref{radialtortoise}) are given as
\begin{align}
\bar{R}(r^*) \sim \left\{ \begin{array}{lr}
  e^{-i(\omega-m\Omega_\text{h}) r^{\ast}},  \hspace{4.4cm} \mbox{ as $r^*\rightarrow -\infty$}, \\
  C_1 e^{-i(\omega-m\Omega_\text{c}) r^{\ast}} + C_2 e^{i(\omega-m\Omega_\text{c}) r^{\ast}}, \hspace{1.1cm} \mbox{ as $r^*\rightarrow +\infty$},
       \end{array} \right.  \label{sol-in}
\end{align}
where only the ingoing wave exists at the outer horizon, $r^*\rightarrow -\infty$. However, at the cosmological horizon, $r^*\rightarrow +\infty$, both ingoing and outgoing waves are considered. Furthermore, the current density of the scalar field is associated with the Wronskian.
\begin{align}
W(\bar{R},\bar{R}^{\ast}) &= \bar{R}\frac{d\bar{R}^{\ast}}{dr_{\ast}} - \bar{R}^{\ast}\frac{d\bar{R}}{dr_{\ast}},
\end{align}
which should be conserved at both asymptotes. Note that, the $\bar{R}^{\ast}$ is complex conjugate of $\bar{R}$. This implies
\begin{align}
\frac{\omega-m\Omega_\text{h}}{\omega-m\Omega_\text{c}}\left(\frac{1}{C_1}\right)^2 &= 1 - \left(\frac{C_2}{C_1}\right)^2,
\end{align}
where ${C_2}/{C_1}$ can be considered as the reflection coefficient. When ${C_2}/{C_1}>1$, the energy of the black hole can be extracted if the frequency of the scalar field satisfies
\begin{align}
m\Omega_c < \omega < m\Omega_h. \label{SRcond}
\end{align}
The energy extraction process is known as superradiance. Moreover, if one set $C_1=0$ in (\ref{sol-in}), thus only outgoing modes are allowed at the cosmological horizon. The boundary condition (\ref{sol-in}) now becomes the quasinormal boundary condition:
\begin{align}
\bar{R}(r^*) \sim \left\{ \begin{array}{lr}
  e^{-i(\omega-m\Omega_\text{h}) r^{\ast}},  \hspace{2.7cm} \mbox{ as $r^*\rightarrow -\infty$} \\
 e^{i(\omega-m\Omega_\text{c}) r^{\ast}}, \hspace{3cm} \mbox{ as $r^*\rightarrow +\infty$}.
       \end{array} \right. \label{QNMsBC}
\end{align}
Herein, the corresponding quasinormal frequency $\omega$ will be discrete and complex-valued. Their imaginary parts will determine the stability of perturbation, with either exponential decay or growth. Therefore in this work, the stabilities of higher-dimensional rotating black holes will be investigated via the study of QNMs.

\subsection{Angular Equation Analysis}

In general, equation (\ref{radialeq}) and (\ref{angulareq}) must be solved simultaneously in order to numerically obtain associated quasinormal frequency. However, for a small rotation limit $(a\omega) \to 0$, the angular eigenvalue $A_{kjm}$ can be analytically obtained by solving (\ref{angulareq}). In this section, based on the higher-dimensional spheroidal harmonics discussed in \cite{Cho:2009wf,Berti:2005gp}, the analytical form of the angular eigenvalue is derived.

We shall first rewrite (\ref{angulareq}) in term of a new variable $x=\cos 2\theta$. The angular equation becomes
\begin{align}
& \left(1-x^2\right)\left(2+a^2\Lambda\left(1+x\right)\right)S''(x) + \left( (d-5)-(d-1)x  -\frac{a^2\Lambda}{2}\left(x+1)(3+(x-1)d+x\right) \right)S'(x) \nonumber \\
& \quad + \frac{1}{2}\left(A_{kjm} - \frac{2j(j+d-5)}{1+x} - \frac{a^2\mu^2}{2}(1+x) + \frac{a^2\omega^2(x-1)}{(2+a^2\Lambda(1+x))} + 2am\omega \right. \nonumber \\
& \quad \left.  + \frac{m^2(2+a^2\Lambda(1+x))}{(x-1)}           \right)S(x)=0, \label{eq:angulareq010}
\end{align}
where $ S' \equiv \frac{dS}{dx}$. Furthermore, we redefine $x=2z-1$, thus (\ref{eq:angulareq010}) becomes
\begin{align}
& 2z(z-1)\left(1+a^2\Lambda z\right)\frac{d^2S}{dz^2} + \left(3-d+(d-1)\left(1-a^2\Lambda\right)z + (d+1)a^2\Lambda z^2\right)\frac{dS}{dz} \nonumber \\
&~~+ \frac{1}{2}\left( a^2\mu^2 z - A_{kjm} + \frac{j(j+d-5)}{z} + \frac{(1-z)a^2\omega^2}{1+a^2\Lambda z} - 2amw + \frac{m^2\left(1+a^2\Lambda z\right)}{(1-z)} \right)S=0. \label{eq:angulareq012}
\end{align}
The ansatz is substituted to obtain Heun's differential equation \cite{Cho:2009wf}
\begin{align}
S(z) &= 2^{\frac{m}{2}}(z-1)^{\frac{m}{2}}(2z)^{\frac{j}{2}}\left(z+\frac{1}{a^2\Lambda}\right)^{\frac{i a\omega}{2\sqrt{a^2\Lambda}}}h(z).
\end{align}
Then, (\ref{eq:angulareq012}) can be put into the Heun's equation.
\begin{align}
\frac{d^2h}{dz^2} + \left(\frac{\gamma}{z} + \frac{\delta}{z-1} + \frac{\epsilon}{z+\frac{1}{a^2\Lambda}}\right)\frac{dh}{dz} + \frac{\left(\alpha\beta z -q\right)}{z(z-1)\left(z+\frac{1}{a^2\Lambda}\right)}h &= 0,
\end{align}
where
\begin{align}
\alpha &= \frac{1}{2}\left(j+m+\frac{i a\omega}{\sqrt{a^2\Lambda}}\right) + \bar{g},\quad
\beta  = \frac{1}{2}\left(j+d-1+m+\frac{i a\omega}{\sqrt{a^2\Lambda}}\right) - \bar{g}, \quad \gamma = \frac{1}{2}\left(2j+d-3\right), \nonumber \\ \delta &= 1+m, \quad
\epsilon = 1 + \frac{ia\omega}{\sqrt{a^2\Lambda}}, \quad
\bar{g} = \frac{1}{4\Lambda}\left((d-1)\Lambda - \sqrt{\Lambda\left((d-1)^2\Lambda-4\mu^2\right)}\right), \nonumber\\
q &= \frac{m\omega}{2a\Lambda}  + \frac{1}{4}\left(j+\frac{ia\omega}{\sqrt{a^2\Lambda}}\right)\left(j+d-3+\frac{ia\omega}{\sqrt{a^2\Lambda}}\right) +  \frac{A_{kjm}-(j+m)(j+d-3+m)}{4a^2\Lambda}.
\end{align}
According to Heun's equation, there are constrained and recurrent relations.
\begin{align}
1+ \alpha+\beta - \gamma - \delta - \epsilon =0, \quad \alpha_0\tilde{a}_1 + \beta_0 \tilde{a}_0 = 0, \quad \alpha_p\tilde{a}_{p+1} + \beta_p \tilde{a}_p + \gamma_p \tilde{a}_{p-1} = 0,\,\,(p=1,2,...),
\end{align}
where
\begin{align}
\alpha_p &= -\frac{(p+1)(p+\tilde{r}-\alpha+1)(p+\tilde{r}-\beta+1)(p+\delta)}{(2p+\tilde{r}+2)(2p+\tilde{r}+1)}, \\
\beta_p &= \frac{p(p+\tilde{r})(\gamma-\delta)\epsilon + \left(p(p+\tilde{r})+\alpha\beta\right)\left(2p(p+\tilde{r})+\gamma(\tilde{r}-1)\right)}{(2p+\tilde{r}+1)(2p+\tilde{r}-1)} - \frac{p(p+\tilde{r})}{a^2\Lambda} - q, \nonumber\\
\gamma_p &= -\frac{(p+\alpha-1)(p+\beta-1)(p+\gamma-1)(p+\tilde{r}-1)}{(2p+\tilde{r}-2)(2p+\tilde{r}-1)},\quad \tilde{r}=j+m+\frac{d-3}{2}.\nonumber
\end{align}
The angular eigenvalue $A_{kjm}$ can be obtained by solving the continued fraction equation \cite{Leaver:1985ax}
\begin{align}
\beta_0 - \frac{\alpha_0\gamma_1}{\beta_1-}\frac{\alpha_1\gamma_2}{\beta_2-}\frac{\alpha_2\gamma_3}{\beta_3-}... &= 0. \label{CFM}
\end{align}
In the absence of rotation, the analytical formula for $A_{kjm}$ is \cite{Berti:2005gp,Kodama:2009rq}
\begin{align}
A_{kjm} &= \left(2k+j+m\right)\left(2k+j+m+d-3\right). \label{norotationA}
\end{align}
If we let $2k=l-(j+m)$, the above formula becomes $A_{kjm}=l(l+d-3)$, which provides the correct form of $A_{kjm}=l(l+1)$ in four dimensions. Furthermore, in higher dimensions, $l$ is constrained to $l \geq j+m$.

In a small $(a\omega)$ range, it is more convenient to use the $k$-th inversion of recurrent relation (\ref{CFM}) \cite{Berti:2005gp,Cho:2009wf} 
\begin{align}
\beta_k - \frac{\alpha_{k-1}\gamma_k}{\beta_{k-1}-}\frac{\alpha_{k-2}\gamma_{k-1}}{\beta_{k-2}-}...\frac{\alpha_0\gamma_1}{\beta_0} &= \frac{\alpha_k\gamma_{k+1}}{\beta_{k+1}-}\frac{\alpha_{k+1}\gamma_{k+2}}{\beta_{k+2}-}... \label{InvCFM}
\end{align}
The angular eigenvalue can be expanded as a power series of the form
\begin{align}
A_{kjm} &= \sum_{\bar{p}=0}^{\infty} f_{\bar{p}} (a\omega)^{\bar{p}}. \label{Aexpand}
\end{align}
By performing series expansion of (\ref{InvCFM}), the coefficients $f_{\bar{p}}$ can be analytically obtained by substituting (\ref{Aexpand}) into (\ref{InvCFM}) and equating the power of $(a\omega)$. The example of these expressions can be found in \cite{Berti:2005gp,Cho:2009wf,Suzuki:1998vy}. An example of an angular eigenvalue computed via the proposed formula is shown in Table~\ref{tab:tab1}. The separation constant $A_{kjm}$ is calculated up to $\mathcal{O}(\bar{c}^2)$ and $\mathcal{O}(\bar{\alpha})$, where $\bar{c}$ and $\bar{\alpha}$ are defined as $a\omega$ and $a^2\Lambda$, respectively. These results are shown to be in good agreement with the small $\bar{c}$ expansion formula given in Ref. \cite{Cho:2009wf}.

\setlength{\tabcolsep}{15pt}
\begin{table}[h]

\centering

\begin{tabular}{|c|c|c|c|}
\hline
\{ $A_{kjm},\mu$ \} & $\bar{c}=0.01,\bar{\alpha}=0.07$ & $\bar{c}=0.05,\bar{\alpha}=0.06$ & $\bar{c}=0.1,\bar{\alpha}=0.05$   \\  \hline 
$\{{A_{010},0}\}$ & \{4.0200,~\textit{4.0200}\}  & \{4.0179,~\textit{4.0178}\} &  \{4.0172,~\textit{4.0170}\}   \\
\hline 
\{${A_{022},0.1}$\} & 28.6170  & 28.3656  & 28.0793    \\
\hline
\end{tabular}
\caption{An example of angular eigenvalue $A_{kjm}$ for small $\bar{c}$ expansion for $d=6$. The results labeled as \textit{italic} font are computed via the formula given in \cite{Cho:2009wf}. }
\label{tab:tab1}
\end{table}

\section{Near Nariai-Type Extremal Limit}\label{sec:nearextremal}

The QNMs can be numerically obtained by solving (\ref{radialeq}) and (\ref{angulareq}). However, in the near Nariai-type extremal limit, $r_h\approx r_c$, the effective potential in the radial equation can be reduced to the P\"oschl-Teller potential \cite{Poschl1933}, where the quasinormal spectrum $\omega$ can be explicitly obtained \cite{Ferrari:1984zz}. Then, the quasinormal frequencies can be analytically obtained at the near Nariai-type extremal limit. The P\"oschl-Teller approximation technique is widely used for investigating the QNMs of near Nariai-type extremal dS black holes or black strings \cite{Cardoso:2003sw,Zhidenko:2003wq,Molina:2003ff,Chang:2005ki,Ponglertsakul:2018smo}. Remark that, to create the near Nariai-type extremal black hole, spacetime parameters such as $a$ and $\Lambda$ must be fine-tuned.

Herein, the analytical form of the quasinormal frequency is derived within the near Nariai-type extremal limit of MP-dS black holes with a single rotation. We start by first noticing that $\Delta_r$ and the surface gravity in (\ref{surfacegravity}) can be rewritten as
\begin{align}\label{deltanear}
\Delta_r &= \frac{\Lambda}{r^{d-5}}(r-r_h)(r_c-r)(r-r_1)...(r-r_{d-3}), \nonumber \\
\kappa_h &= \frac{\Lambda}{2(r_h^2+a^2)r_h^{d-5}}(r_c-r_h)(r_h-r_1)...(r_h-r_{d-3}).
\end{align}
In the near Nariai-type extremal limit, they become
\begin{align}
\Delta_r \approx (d-1)(r_c-r)(r-r_h)r_h^2\Lambda,\quad \kappa_h \approx \frac{(d-1)(r_c-r_h)r_h^2\Lambda}{2(r_h^2+a^2)}.
\end{align}
The tortoise coordinate can be integrated
\begin{align}
r^{\ast} &= \frac{\left(r_h^2+a^2\right)}{(d-1)r_h^2\Lambda}\int \frac{1}{\left(r-r_h\right)\left(r_c-r\right)}dr, \nonumber \\
&= \frac{\left(r_h^2+a^2\right)}{(d-1)(r_c-r_h)r_h^2\Lambda}\left(\ln(r-r_h) - \ln(r_c-r)\right). 
\end{align}
Then, the radial coordinate is rewritten in term of tortoise coordinate
\begin{align}
r &= \frac{r_h + r_c e^{\tilde{K}r^\ast}}{1+e^{\tilde{K}r^\ast}},
\end{align}
where we define $\tilde{K}\equiv 2\kappa_h$. Therefore, the metric $\Delta_r$ in the near Nariai-type extremal limit is expressed as a function of $r^{\ast}$
\begin{align}
\Delta_r &\sim \frac{(d-1)(r_c-r_h)^2 r_h^2\Lambda}{4\cosh^2(\tilde{K}r^\ast/2)}.
\end{align}
This reduces the radial equation (\ref{radialtortoise}).
\begin{align}\label{PTeq}
\frac{d^2 \bar{R}}{{dr^\ast}^2} + \left[\left(\omega-m\Omega_h\right)^2 - \frac{V_0}{\cosh^2\left(\tilde{K}r^{\ast}/2\right)}  \right]\bar{R} = 0,\,  
\end{align}
where
\begin{align}
V_0 &= \frac{(d-1)(r_c-r_h)^2r_h^2\Lambda}{4\left(r_h^2 + a^2\right)^2}\left[\frac{j(j+d-5)a^2}{r_h^2} + \mu^2r_h^2 + A_{kjm}\right].
\end{align}
This is the well-known P\"oschl-Teller potential \cite{Poschl1933}. With the quasinormal boundary condition (\ref{QNMsBC}), the near-extremal radial wave equation can be solved analytically \cite{Ferrari:1984zz}. The associated quasinormal frequency $\omega$ is given by
\begin{align}
\omega_n &= m\Omega_h + \kappa_h \left[\sqrt{\frac{4V_0}{\tilde{K}^2}- \frac{1}{4}} - i\left(n+\frac{1}{2}\right), \right].
\end{align}
where $n=0,1,2,...$ is the overtone number. The real part of the quasinormal frequency depends on the scalar field contents, whereas the imaginary part is only related to the surface gravity of the black hole's horizon. Similarly, the proposed formula is derived in the non-minimally coupled scalar field with Einstein's gravity \cite{Gwak:2019ttv}.

\begin{figure}[h]
        \centering
        \includegraphics[width=0.45\textwidth]{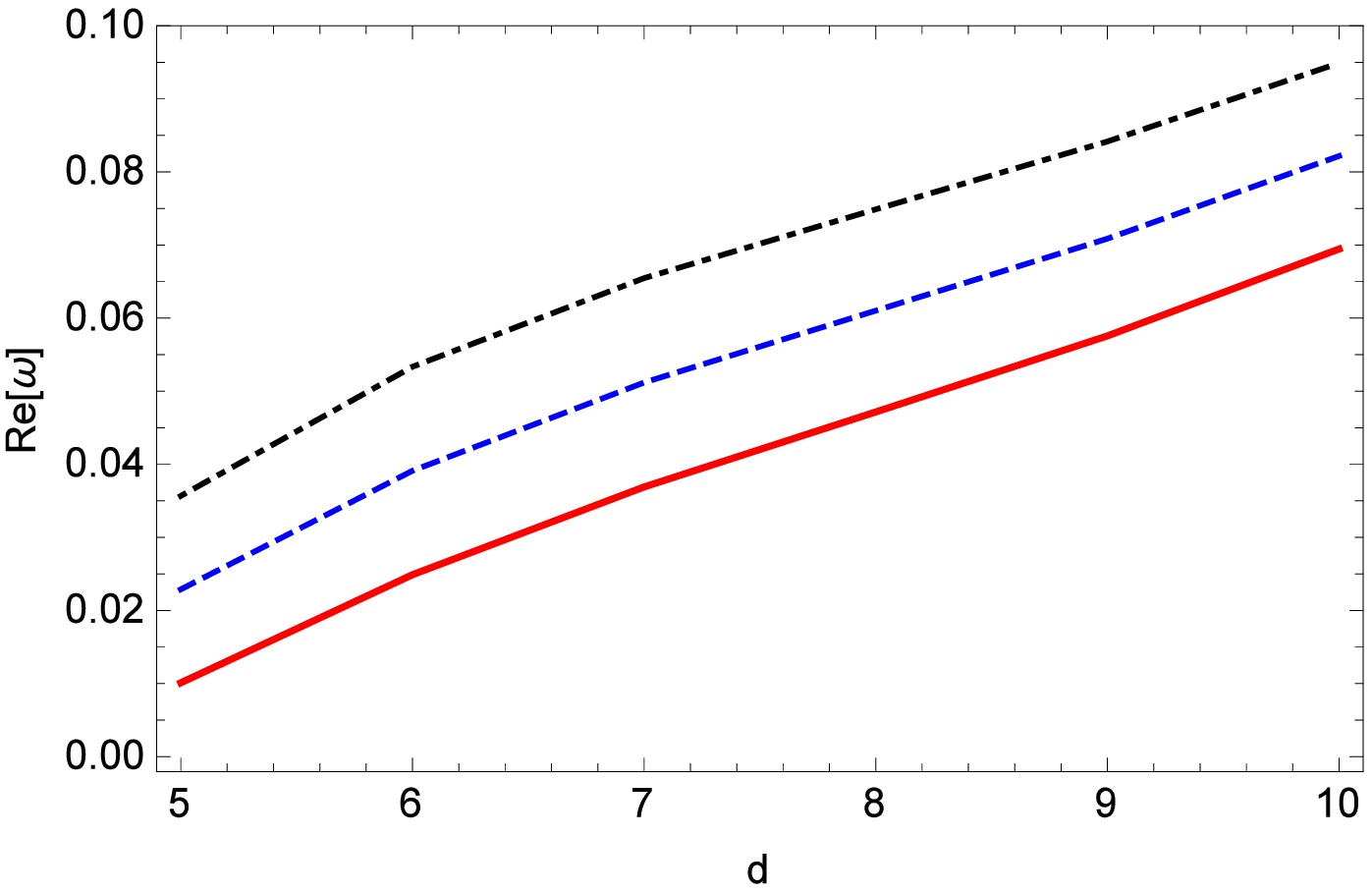}  
        \includegraphics[width=0.45\textwidth]{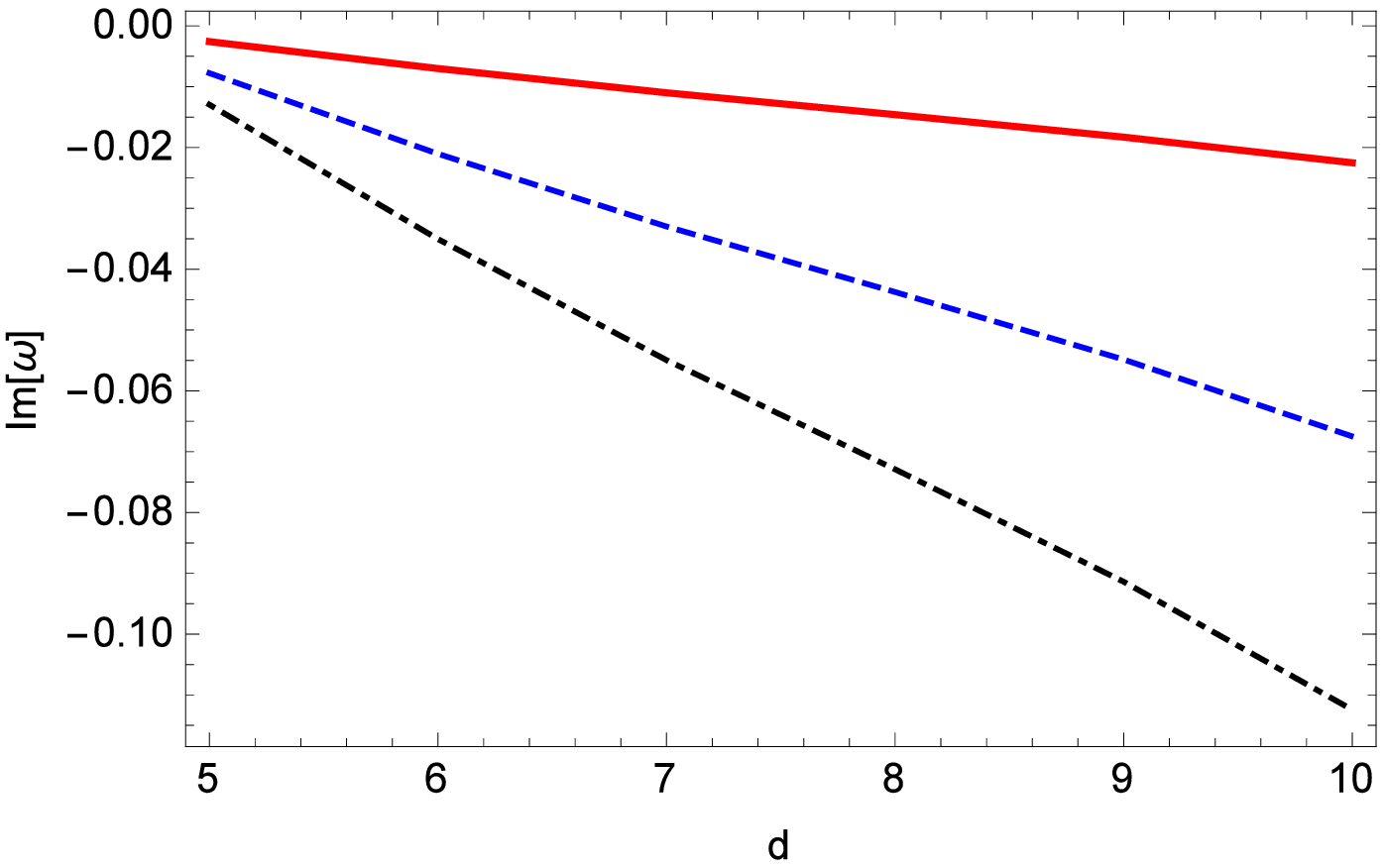}  
        \caption{The quasinormal frequencies (real and imaginary parts) $\omega$ plot against the number of spacetime dimensions $d$. For near extremal black holes with $M=1,a=0.1,\mu=0.02,k=0$. (Red, solid): $j=2,m=n=0$. (Blue, dashed): $j=1,m=n=1$. (Black, dotdashed): $j=0,m=n=2$} \label{PTplots}
\end{figure}

The quasinormal frequencies of near Nariai-type extremal MP-dS black holes with a single rotation are shown in Fig~\ref{PTplots}. In this figure, the black hole mass and spin parameter are fixed at $1$ and $0.1$, respectively, for $d=5-10$. Thus, the cosmological constant will be properly adjusted to create a near-extremal scenario in each dimension. From the black hole phase spaces in Fig~\ref{phasespace}, it can be expected that the value of $\Lambda$, for which the near Nariai-type extremal limit is reached, will increase as $d$ increases for a fixed $a$ value. For $d=5-10$, the $\Lambda$ used in these plots ranges between $0.125$ and $0.415$. In addition, the parameters are chosen such that $(r_c-r_h)$ is always less than $0.03$. From the plots, it can be observed that Re$(\omega)$ increases, whereas Im$(\omega)$ decreases with the number of spacetime dimensions. Thus, the perturbation modes decay faster in higher dimensions. Moreover, the scalar wave oscillates more and decays faster as the azimuthal number $m$ and overtone number $n$ increase. These near Nariai-type extremal frequencies do not satisfy the superradiant condition (\ref{SRcond}). Therefore, these modes do not absorb rotational energy from the black holes.

Although the formula for the angular eigenvalue $A_{kjm}$ is present, an exact solution of the radial wave equation (\ref{radialeq}) is not possible beyond the near-extremal regime. Thus, another two conventional methods, WKB and AIM, are employed to overcome this problem semi-analytically and numerically. These approaches will be discussed in the following sections.

\section{WKB method}\label{sec:wkb}

Schutz and Will \cite{Schutz:1985zz} firstly applied the WKB approximation to study the black hole scattering problem. It determines the complex frequencies of black holes. According to the WKB approximation, the quasinormal frequency is obtained by analyzing the peak of an effective potential in the radial equation \cite{Konoplya:2011qq}. The approximation method was extended to the third order by Iyer \cite{Iyer:1986np} and later to the sixth-order by Konoplya \cite{Konoplya:2003ii}. Because it is semi-analytical, the WKB method has been used in several studies on black hole QNMs \cite{Zhang:2003vb,Cornell:2005ux,Liu:2008mj,Konoplya:2009ig}. In this section, the QNMs of higher-dimensional MP-dS black holes will be computed by the WKB approximation. 

In general, the radial wave equation can be taken in the form
\begin{align}
\frac{d^2 \bar{R}}{dr_{\ast}^2} + Q(r_{\ast})\bar{R} &= 0,\quad Q\equiv \omega^2 - V(r_{\ast}). \label{radialWKB}
\end{align}
In the non-rotating case, the effective potential $V$ is independent of a complex frequency $\omega$. However, with rotation, $V$ is a function of $\omega$ and $A_{kjm}$. Then, evaluating the peak of the potential becomes non-trivial. To overcome this issue, the series expansion method proposed in \cite{Seidel:1989bp} to investigate the WKB method on the Kerr black hole is utilized. To solve the radial equation, $A_{kjm}$ and $V$ can be expanded as a power series in $(a\omega)$ in a slowly rotating limit. This can be done by expanding $r_{\max}$, the location of the maximum of $V(r_{\max})$.
\begin{align}
r_{max} &= \sum_{i=0}^{6} r_i (a\omega)^i \equiv r_0 + \sigma. \label{rexpand}
\end{align}
The $r_0$ is the location where the effective potential $V$ attains its maximum value in the zero angular momentum case.
To find the coefficients $r_i$, $V'(r_{max})$ is expanded as follows:
\begin{align}
V'(r_0+\sigma) &\approx V'(r_0) + \sigma V''(r_0) +\frac{\sigma^2}{2}V'''(r_0) + \frac{\sigma^3}{6}V^{(4)}(r_0)+ \frac{\sigma^4}{24}V^{(5)}(r_0) + \frac{\sigma^5}{120}V^{(6)}(r_0) \nonumber \\
&+ \frac{\sigma^6}{720}V^{(7)}(r_0)=0. \label{Vprimeexpand}
\end{align}
By substituting (\ref{rexpand}) into (\ref{Vprimeexpand}), the coefficients $r_i$ can be computed perturbatively, order by order, up to $(a\omega)^6$. Then, $r_0....r_6$ quantities can be expressed in term of $a$ and $\omega$. Now the $r_{max}$ can be constructed via (\ref{rexpand}). Therefore, we can evaluate the derivative potential $V$ at $r_{max}$. The sixth-order WKB formula for the quasinormal frequency is given by \cite{Konoplya:2003ii}.
\begin{align}
\frac{i Q_{max}}{\sqrt{2Q''_{max}}} - \Lambda_2 - \Lambda_3 - \Lambda_4 - \Lambda_5 - \Lambda_6 &= n+\frac{1}{2}, \label{WKB}
\end{align}
where the correction terms $\Lambda_2$ and $\Lambda_3$ can be found in \cite{Iyer:1986np}, and $\Lambda_4,\Lambda_5$, and $\Lambda_6$ are defined in \cite{Konoplya:2003ii}. $Q''_{max}$ denotes the second derivative of $Q$ with respect to the coordinate $r_{\ast}$, and it must be evaluated at $r_{max}$. The correction terms $\Lambda_{2-6}$ are also related to a higher derivative term of $Q$. At the sixth-order WKB, the highest derivative term is $Q^{(12)}_{max}$.
\setlength{\tabcolsep}{15pt}
\begin{table}[h]

\centering

\begin{tabular}{|c|c|c|}
\hline
$\Lambda$  & WKB 3rd order & WKB 6th order    \\  \hline

0 & 0.48321 $-$ 0.09680$i$  & 0.48364 $-$ 0.09677$i$ \\
 
\hline

0.0067 & 0.43426 $-$ 0.08862$i$  & 0.43461 $-$ 0.08858$i$ \\
 
\hline
 
0.02 & 0.31987 $-$ 0.06687$i$ & 0.32002 $-$ 0.06685$i$   \\
 
\hline

0.03 & 0.20290 $-$ 0.04258$i$ & 0.20295 $-$ 0.04256$i$   \\
 
\hline

0.0367 & 0.04641 $-$ 0.01075$i$ & 0.04367 $-$ 0.00964$i$ \\

\hline

%
 
\end{tabular}
\caption{Comparison of the 3rd order WKB and the 6th order WKB for four dimensional non-rotating black holes with $M=1,l=2,n=0,\mu=0$.}
\label{tab:tab2}
\end{table}
The validity of (\ref{WKB}) is verified by setting the non-rotating case, $a=0$, in (\ref{radialWKB}). Note that in this case, the separation constant becomes $A_{kjm}=l(l+d-3)$. WKB formula (\ref{WKB}) reduces to a simple polynomial equation in $\omega$. In Table~\ref{tab:tab2}, the quasinormal frequencies of the massless scalar field are computed in four-dimensional Schwarzschild-dS spacetime. These results are in good agreement with those reported in \cite{Zhidenko:2003wq}.

Solving (\ref{WKB}) is more complicated for a rotating case than for non-rotating case. This is because $V$ and $A_{kjm}$ are $\omega$-dependent, and hence, each correction term contains a higher degree of polynomials $\omega$. Multiple roots of $\omega$ are obtained when solving (\ref{WKB}). We begin by calculating the quasinormal frequencies in the Schwarzschild-dS limit. Then, the spin parameter gradually increases, and the WKB equation is solved about each parameter. The root $\omega$ can be found using the $\omega$ from the previous step as an initial value. This process is repeated until a desired spin parameter value is reached.
\setlength{\tabcolsep}{15pt}
\begin{table}[h]

\centering

\begin{tabular}{|c|c|c|c|c|}
\hline
\{$d,~a,~\Lambda$\}  & WKB 3rd order & PT & $m\Omega_c$ & $m\Omega_h$  \\  \hline

\{4,~0.05,~0.03703\} & 0.0217 $-$ 0.0026$i$  & 0.0222 $-$ 0.0026$i$ & 0.0035 & 0.0039\\
 
\hline

\{5,~0.10,~0.12490\} & 0.0899 $-$ 0.0144$i$   & 0.0936 $-$ 0.0143$i$ & 0.0111 & 0.0140 \\
 
\hline
 
\{6,~0.15,~0.20519\} & 0.1560 $-$ 0.0282$i$ & 0.1634 $-$ 0.0280$i$ & 0.0174 & 0.0240  \\
 
\hline

\{7,~0.20,~0.27210\} & 0.2413 $-$ 0.0478$i$ & 0.2542 $-$ 0.0470$i$ & 0.0215 & 0.0337  \\
 
\hline

\end{tabular}
\caption{Comparison between the quasinormal frequencies computed via 3rd order WKB and P\"oschl-Teller formula for near extremal MP-dS black holes in $d=4-7$. The background parameters are chosen as $M=1,k=0,j=2,m=1,n=0,\mu=0.01$.}
\label{tab:tab3}
\end{table}
The comparative results of the respective WKB and P\"oschl-Teller formulas are shown in Table~\ref{tab:tab3}. Note that in the near-extremal regime, only the third-order WKB is applicable. It is found that as $r_h$ approaches $r_c$, the higher derivative term of $Q_{max}$ becomes zero. Thus, the higher correction terms $\Lambda_{4-6}$ diverge. For this reason, when comparing the WKB and P\"oschl-Teller formulas, the parameters are chosen such that $0.05<(r_c-r_h)<0.15$. Despite the lack of a result comparable to the sixth-order WKB, the quasinormal frequencies from these two methods are in good agreement. It is noted that as $d$ and $a$ increase, the difference becomes more significant. Herein, it is concluded that these modes do not reside in the superradiant regime.

\section{Asymptotic Iteration Method}\label{sec:AIM}

A detailed study on QNMs can be conducted using a numerical approach. Herein, AIM is implemented. This method was firstly used for an eigenvalue problem \cite{AIM:2003}. Later, it was applied to calculate the quasinormal frequencies of a Schwarzschild black hole in asymptotically flat and dS spacetimes \cite{Cho:2009cj}. Then, it was generalized to higher-dimensional cases for the spheroidal harmonic of Kerr-(A)dS and doubly rotating black holes (two equal angular momenta) \cite{Cho:2009wf,Cho:2011yp}. In recent times, this method has been widely used to investigate QNMs in alternative theories of gravity \cite{Ponglertsakul:2018smo,Prasia:2016fcc,Burikham:2017gdm,Tangphati:2018jdx,Burikham:2019fza,Zangeneh:2017rhc}.

Compared to Leaver's continued fraction method (CFM) \cite{Leaver:1985ax}, which is more frequently used to determine QNMs, the AIM has certain advantages. The CFM generally involves a lengthy and complicated calculation to obtain recurrence relation coefficients, which are primary factors in the method. Moreover, a Gaussian elimination process may be required to reduce a recurrence relation to a three-term recurrence relation. The AIM does not require these processes. Therefore, given the complexity of the master equation herein, those complicated steps, which are prone to error, can be bypassed using the AIM. In this study, the radial (\ref{radialeq}) and angular (\ref{angulareq}) wave equations will be solved using the AIM to obtain the quasinormal frequency of the MP-dS black hole with a single rotation.

\subsection{AIM for the Radial Equation}

For applying AIM to obtain the quasinormal frequency, a new variable $\xi=1/r$ is introduced to the radial equation (\ref{radialeq})
\begin{align}
\frac{d^2R}{d\xi^2} + \left[\frac{\Delta'_{\xi}}{\Delta_{\xi}} - \frac{d-6}{\xi}\right]\frac{dR}{d\xi} + \frac{1}{\xi^8 \Delta_{\xi}^2}\left[\left(\omega+a^2\omega\xi^2 - am\left(\xi^2 - \Lambda\right)\right)^2 \right. \nonumber \\
\left. - \xi^2\Delta_{\xi}\left(A_{kjm}\xi^2 + a^2j(j+d-5)\xi^4+\mu^2\right)\right]R &= 0, \label{radialinX}
\end{align}
where $\Delta'_{\xi}\equiv \frac{d\Delta_{\xi}}{d\xi}$. The domain of $\xi$ varies from $\xi_c=1/r_c$ to $\xi_h=1/r_h$. 
In terms of the new variable, it is useful to consider
\begin{align}
e^{i\omega r_{\ast}} = \prod_{i=h}^{d-3} \left(\xi-\xi_i\right)^{\frac{i\omega}{2\kappa_i}}, \quad 
\kappa_i = (-1)^{d-1}\frac{\Lambda}{2}\left[\frac{\prod\limits_{i \neq j,j=h }^{d-3}(\xi_i-\xi_j)}{\prod\limits_{i=h}^{d-3}\xi_i(1+a^2\xi_i)}\right],
\end{align} 
where the indices $i,j$ range from $\{h,c,1,...,d-3\}$. The tortoise coordinate can be rewritten as 
\begin{align*}
dr_{\ast} &= -\frac{\xi^{-2}+a^2}{\xi^2\Delta_{\xi}}d\xi,\quad \Delta_{\xi} = -\frac{\Lambda}{\xi^4}\frac{(\xi_h-\xi)(\xi_c-\xi)(\xi_1-\xi)...(\xi_{d-3}-\xi)}{\left(\xi_h\xi_c\xi_1...\xi_{d-3}\right)}.
\end{align*}
To scale out the divergence at the cosmological horizon, 
\begin{align}
R &\equiv e^{i\omega r_{\ast}}u(\xi).
\end{align}
The wave equation (\ref{radialinX}) can now be stated as
\begin{align}
\frac{d^2u}{d\xi^2} + \left[\frac{\Delta'_{\xi}}{\Delta_{\xi}} - \frac{d-6}{\xi} - \frac{2i\omega\left(1+a^2\xi^2\right)}{\xi^4\Delta_{\xi}}\right]\frac{du}{d\xi}  + \left[\frac{\mathcal{W}_0 + \mathcal{W}_1\omega}{\xi^8\Delta_{\xi}^2}\right]u   &= 0,  \label{radialinU}
\end{align}
where
\begin{align}
\mathcal{W}_0 &= a^2m^2\left(\xi^2 - \Lambda\right)^2 - \Delta_{\xi}\xi^2\left(A_{kjm}\xi^2+j(j+d-5)a^2\xi^4+\mu^2 \right),  \\
\mathcal{W}_1 &= -\frac{i(d-2)}{\xi}\left(2M\xi^{d-1}+\Lambda-\xi^2\right) - 2ma\left(\xi^2-\Lambda\right)  -2i\left((d-4)M\xi^d  \right. 
\nonumber \\
& \quad \left. -(d-3)(\xi^2-\Lambda)\xi \right)  a^2 - 2m\xi^2\left(\xi^2-\Lambda\right) a^3 + i(d-4)(\xi^2-\Lambda)\xi^3a^4.
\end{align}
At the outer horizon, the divergence is scaled out by setting
\begin{align}
u \equiv (\xi-\xi_h)^{-\frac{i\omega}{\kappa_h}}\chi(\xi).
\end{align}
Then, the wave equation (\ref{radialinU}) takes the form
\begin{align}
\frac{d^2\chi}{d\xi^2} &= \lambda_0 (\xi,\omega)\frac{d\chi}{d\xi} + s_0(\xi,\omega)\chi, \label{radialAIM}
\end{align}
where
\begin{align}
\lambda_0 &= \frac{d-6}{\xi} + \frac{2i\omega}{(\xi-\xi_h)\kappa_h} + \frac{2i\omega\left(1+a^2\xi^2\right)}{\xi^4\Delta_{\xi}} - \frac{\Delta'_{\xi}}{\Delta_{\xi}}, \quad s_0 = \mathcal{\bar{W}}_0 + \mathcal{\bar{W}}_1 \omega + \mathcal{\bar{W}}_2 \omega^2, \nonumber \\
\mathcal{\bar{W}}_0 &= -\frac{\mathcal{W}_0}{\xi^8\Delta_{\xi}^2}, \nonumber\\
\mathcal{\bar{W}}_1 &= -\frac{i}{\xi^{10} \Delta_{\xi}^2} \Biggl[ -2M\xi^d \left(d-2+(d-4)a^2\xi^2\right) + \xi\left(1+a^2\xi^2\right) \nonumber \\
&\times  \left(d-2 + 2 i a m \xi + (d-4)a^2\xi^2\right)\left(\xi^2-\Lambda\right)  \nonumber \\
& +  \frac{\xi^9\Delta_{\xi}\left( \{(d-5)\xi - (d-6)\xi_h\}\Delta_{\xi} - \xi\{\xi-\xi_h\}\Delta'_{\xi}           \right)}{(\xi-\xi_h)^2\kappa_h}
        \Biggr], \nonumber\\
\mathcal{\bar{W}}_2 &= \frac{2(1+a^2\xi^2)(\xi-\xi_h)\kappa_h + \xi^4\Delta_\xi}{\xi^4(\xi-\xi_h)^2\kappa_h^2\Delta_{\xi}}.\nonumber
\end{align}
Equation\,(\ref{radialAIM}), coefficients $\lambda_0$ and $s_0$ are the core of the AIM. The coefficients $\lambda_0$ and $s_0$ will be fed into the AIM numerical routine. According to the algorithm of AIM, by differentiating $n$ times with respect to $\xi$ on (\ref{radialAIM}), it is seen that \cite{Cho:2009cj}
\begin{align}
\chi^{(n)} = \lambda_{n-2}\chi' + s_{n-2}\chi,
\end{align}
where the recurrent formulas for the coefficients $\lambda_{n-2}$ and $s_{n-2}$ are defined by
\begin{align}
\lambda_n &= \lambda'_{n-1} + \lambda_{n-1}\lambda_0 + s_{n-1}, \label{iter1} \\
s_n  &= s'_{n-1} + s_0\lambda_{n-1}. \label{iter2}
\end{align}
Herein, the prime denotes the derivative with respect to $\xi$. For a sufficiently large $n$, the asymptotic aspect of AIM is
\begin{align}
\frac{s_n}{\lambda_{n}} = \frac{s_{n-1}}{\lambda_{n-1}}\equiv P,
\end{align}
where $P$ is a constant. The quantization condition can be solved for the quasinormal frequency \cite{Cho:2009cj} 
\begin{align}
\lambda_n (\xi)s_{n-1}(\xi) &= \lambda_{n-1}(\xi)s_n(\xi). \label{quantcond}
\end{align}
To determine the eigenvalue spectrum, each coefficient $\lambda_n,s_n$ will be constructed in terms of its previous iteration via (\ref{iter1}) and (\ref{iter2}). At each $n$-th iteration, the derivative of $\lambda_{n-1},s_{n-1}$ must be calculated. This is the main drawback of the AIM because it significantly increases the computational time and also affects the numerical precision \cite{Cho:2009cj,Prasia:2016fcc}. To overcome this problem, an improved version of AIM was proposed in \cite{Cho:2009cj}. The derivative at each iteration can be reduced by expanding $\lambda_n,s_n$ at a specific point $\bar{\xi}$.
\begin{align}
\lambda_n(\bar{\xi}) &= \sum_{i=0}^{\infty} c^{i}_n (\xi-\bar{\xi})^i, \\
s_n(\bar{\xi}) &= \sum_{i=0}^{\infty} d^{i}_n (\xi-\bar{\xi})^i,
\end{align}
where the $i$-th coefficients of $\lambda_n$ and $s_n$ are denoted by $c^{i}_n$ and $d^{i}_n$, respectively. Equations\,(\ref{iter1}) and (\ref{iter2}) can now be expressed as
\begin{align}
c^{i}_n = (i+1)c^{i+1}_{n-1} + d^i_{n-1} + \sum_{k=0}^{i}c^{k}_0 c^{i-k}_{n-1},\quad 
d^{i}_n &= (i+1)d^{i+1}_{n-1} + \sum_{k=0}^{i}d^{k}_0 c^{i-k}_{n-1}. \label{iter2a}
\end{align}
The quantization condition written in terms of $c^{i}_n$ and $d^{i}_n$ is obtained as follow:
\begin{align}
d^{0}_{n}c^{0}_{n-1} - d^{0}_{n-1}c^{0}_{n} &= 0. \label{quantcond1}
\end{align}
It can be observed that the final recurrence relation does not require the derivative operator. The quasinormal frequency is obtained by solving this simple recurrence relation. Clearly, the improved AIM relies on some expansion point $\bar{\xi}$. It is found that using a different expansion point can either improve or hinder the speed of the convergence \cite{Cho:2009wf}. In addition, the AIM appears to converge the fastest when $\bar{\xi}$ is chosen to be at the maximum point of an effective potential \cite{Barakat_2005,Barakat_2006}. However, in higher-dimensional rotating spacetime, the maximum point of the effective potential cannot be explicitly determined. Therefore, throughout this study, $\bar{\xi}$ is defined as a middle point between $\xi_c$ and $\xi_h$, unless stated otherwise.

\subsection{AIM for Angular Equation}

Now we will implement the AIM technique to the angular equation (\ref{angulareq}), the new variable $x=\cos\theta$ is introduced. Then, the angular equation becomes
\begin{align}
&\left(1-x^2\right)\left(1+a^2\Lambda x^2\right)\frac{d^2S}{dx^2} +\Biggl( \frac{(d-4)(1-x^2)-x^2}{x} + a^2\Lambda(d-2-(d-1)x^2)x \\
& \quad - \left(1+a^2\Lambda x^2\right)x \Biggr) \frac{dS}{dx} + \Biggl ( A_{kjm} + 2 a m \omega  - a^2\mu^2x^2 - \frac{j(j+d-5)}{x^2} - \frac{a^2\omega^2(1-x^2)}{1+a^2\Lambda x^2}  \nonumber \\
& \quad -\frac{m^2\left(1+a^2\Lambda x^2\right)}{1-x^2} \Biggr)S=0, \nonumber \label{angularinX}
\end{align}
which can be transformed into the suitable form of the AIM by setting the angular ansatz as follows \cite{Cho:2009wf}:
\begin{align}
S &= (1-x^2)^{\frac{m}{2}}w(x).
\end{align}
The angular equation in Eq.\,(\ref{angularinX}) can be rewritten as
\begin{align}
\frac{d^2w}{dx^2} &= \bar{\lambda}_0 \frac{dw}{dx} + \bar{s}_0 w,
\end{align}
where
\begin{align}\label{eq:coefficients0102}
\bar{\lambda}_0 &= \frac{d+2}{x}\left(\frac{1+m x^2}{x^2-1} - \frac{1}{1+a^2\Lambda x^2}\right), \quad \bar{s}_0 = \bar{\mathcal{V}}_0 + \bar{\mathcal{V}}_1\omega + \bar{\mathcal{V}}_2\omega^2, \\
\bar{\mathcal{V}}_0 &= \frac{j^2 + j(d-5) + x^2 \left(-A_{kjm} + m^2\left(1+a^2\Lambda x^2\right) + m \left(d-3+(d-1)a^2\Lambda x^2\right) + a^2\mu^2x^2\right)}{x^2(1-x^2)(1+a^2\Lambda x^2)}, \nonumber\\
\bar{\mathcal{V}}_1 &= -\frac{2am}{(1-x^2)(1+a^2\Lambda x^2)}, \quad \bar{\mathcal{V}}_2 = \frac{a^2}{(1+a^2\Lambda x^2)^2}.\nonumber
\end{align}
These coefficients will be fed into the numerical routine of AIM as explained in the previous subsection. Then, the eigenvalue $A_{jkm}$ computed via the small $\bar{c}$ expansion formula, and the AIM is compared in Table~\ref{tab:tab3a}. The results of the two methods are shown to be in good agreement. Note that, throughout this work, the AIM expansion point for solving the angular equation is fixed at $x=0.5$.
\begin{table}[h]

\centering

\begin{tabular}{|c|c|c|c|c|}
\hline
\{ $\bar{c}, \bar{\alpha}$ \} & \{0.01, 0.05 \}  & \{0.05, 0.1 \}  &  \{0.1, 0.15 \} & \{0.15, 0.2 \}  \\  \hline

Formula  & 15.2100   & 15.3614  & 15.4957   & 15.6329  \\
 
\hline
 
AIM &  15.2079  & 15.3529  & 15.4769   & 15.5997 \\
 
\hline

%
 
\end{tabular}
\caption{Comparison of the small $\bar{c}$ formula and the AIM of the angular eigenvalue $A_{021}$ for the MP-dS black hole in five dimensions with $M=1,\mu=0.05$. }
\label{tab:tab3a}
\end{table}

\section{Results}\label{sec:results}
In this section, the quasinormal frequency of the massive scalar perturbation on the MP-dS black hole with a single rotation is numerically calculated in both small and large $a$ regimes. Our numerical scheme is based on the improved AIM discussed in the previous section. In slowly rotating regime, the QNMs will be computed by solving the radial equation with the analytic formula for the angular eigenvalue $A_{kjm}$.

Beyond the slowly rotating limit, the numerical procedure is as follow. The QNMs and eigenvalue $(\omega,A_{kjm})$ are calculated in either the Schwarzschild or small rotation limit. At this point, the spin parameter would increase slightly. The eigenvalue $A_{kjm}$ from the previous step is fed into the AIM routine of the radial equation for which the new QNMs are obtained as $\omega'$. 
The new eigenvalue $A'_{kjm}$ is determined by solving the angular equation with the new QNMs $\omega'$. Repeating this procedure (with the same value of $a$), the corrected roots of $(\omega',A'_{kjm})$ are found by searching for the value closest to $(\omega, A_{kjm})$. The converged result is obtained following the iteration of this process. The iteration is stopped when the desired numerical precision is reached. When this process is completed, the spin parameter $a$ is increased by a small value, and the numerical process is then repeated. In this manner, the QNMs in the large $a$ regime are obtained. Herein, the number of iterations is set to seven, for which the majority of the cases explored herein converge perfectly. However, in certain cases with large $a$ values, the number of iterations has to be increased to attain the desired precision.

\subsection{Slowly Rotating Regime}

In Table~\ref{tab:tab4}, the QNMs of near Nariai-type extremal MP-dS black holes in $d=5-9$ computed via the AIM, third-order WKB, and P\"oschl-Teller formula within a small rotation limit are shown. It is seen that as $d$ increases, the real part of $\omega$ increases, whereas the imaginary part decreases. This trend agrees with those observed in Fig~\ref{PTplots} and Table~\ref{tab:tab3}. Interestingly, the difference of Re$(\omega)$ between these three approaches is larger than that in their Im$(\omega)$. The imaginary parts of the quasinormal frequencies appear to be in better agreement across these approaches. Additionally, AIM and WKB are more consistent with one another in terms of the real part, whereas AIM and P\"oschl-Teller concur better in terms of the imaginary part. The result of the P\"oschl-Teller method may be improved by reducing the difference between each horizon. However, the WKB would break down with such a limit. Table~\ref{tab:tab5} presents the comparison between the AIM and sixth-order WKB in non-near Nariai-type extremal cases. As expected, these two methods agree with the sixth-order WKB than the third-order WKB. However, the difference becomes more significant when $d$ increases.
\setlength{\tabcolsep}{15pt}
\begin{table}[h]

\centering

\begin{tabular}{|c|c|c|c|}
\hline
\{$d,~\Lambda$\}  & AIM  & WKB 3rd order  & PT \\  \hline

\{5,~0.1247\} & 0.0491 $-$ 0.0123$i$  & 0.0489 $-$ 0.0123$i$ & 0.0503 $-$ 0.0123$i$ \\
 
\hline

\{6,~0.2049\} & 0.0545 $-$ 0.0150$i$   & 0.0542 $-$ 0.0151$i$ & 0.0553 $-$ 0.0150$i$ \\
 
\hline
 
\{7,~0.2710\} & 0.1152 $-$ 0.0343$i$ & 0.1150 $-$ 0.0343$i$  & 0.1184 $-$ 0.0342$i$  \\
 
\hline

\{8,~0.3260\} & 0.1605 $-$ 0.0497$i$ & 0.1602 $-$ 0.0499$i$ & 0.1654 $-$ 0.0496$i$  \\
 
\hline

\{9,~0.3740\} & 0.1224 $-$ 0.0388$i$ & 0.1221 $-$ 0.0390$i$ & 0.1247 $-$ 0.0388$i$  \\

\hline
 
\end{tabular}
\caption{Comparison between the quasinormal frequencies computed via the AIM, 3rd order WKB and P\"oschl-Teller formula for near extremal MP-dS black holes in $d=5-9$. The background parameters are chosen as $M=1,a=0.01,k=0,j=1,m=1,n=0,\mu=0$.}
\label{tab:tab4}
\end{table}

\setlength{\tabcolsep}{15pt}
\begin{table}[h]

\centering

\begin{tabular}{|c|c|c|}
\hline
$d$ & AIM  & WKB 6th order \\  \hline

5 & 0.2667 $-$ 0.1222$i$  & 0.2663 $-$ 0.1230$i$  \\
 
\hline

6 & 0.7263 $-$ 0.3199$i$   & 0.7252 $-$ 0.3222$i$  \\
 
\hline
 
7 & 1.1364 $-$ 0.4714$i$ & 1.1309 $-$ 0.4804$i$    \\
 
\hline

8 & 1.5379 $-$ 0.6041$i$ & 1.5215 $-$ 0.6272$i$   \\
 
\hline

9 & 1.9378 $-$ 0.7242$i$ & 1.9005 $-$ 0.7716$i$   \\

\hline
 
\end{tabular}
\caption{Comparison between the quasinormal frequencies computed via the AIM and 6th order of WKB for MP-dS black holes in $d=5-9$. The background parameters are chosen as $M=1,a=0.1,\Lambda=0.1,k=0,j=1,m=0,n=0,\mu=0$.}
\label{tab:tab5}
\end{table}

\setlength{\tabcolsep}{15pt}
\begin{table}[h]

\centering

\begin{tabular}{|c|c|c|c|c|}
\hline
$a$ & $j=0$  & $j=1$ & $m\Omega_c$ & $m\Omega_h$ \\  \hline

0 & 0.2655 $-$ 0.1215$i$  & 0.4424 $-$ 0.1152$i$ & 0 & 0  \\
 
\hline

0.1 & 0.2860 $-$ 0.1145$i$   & 0.4546 $-$ 0.1152$i$ & 0.0038 & 0.0263 \\
 
\hline
 
0.2 & 0.3125 $-$ 0.1110$i$ & 0.4725 $-$ 0.1143$i$ & 0.0075 & 0.0533   \\
 
\hline

0.3 & 0.3422 $-$ 0.1096$i$ & 0.4971 $-$ 0.1132$i$ & 0.0110 & 0.0817   \\
 
\hline

0.4 & 0.3744 $-$ 0.1093$i$ & 0.5286 $-$ 0.1128$i$ & 0.0143 & 0.1122  \\

\hline
 
\end{tabular}
\caption{The QNMs for scalar perturbation of five dimensional rotating black hole in dS spacetime computed via AIM. With $M=1,\Lambda=0.1,k=0,m=1,\mu=0.05$.}
\label{tab:tab6}
\end{table}

Table~\ref{tab:tab6}--\ref{tab:tab8} the quasinormal frequencies of slowly rotating black holes in five, six, and seven dimensions are displayed. In each table, the QNMs are compared between $j=0$ and $j=1$ modes. As the spin parameter $a$ increases, the real and imaginary parts of $\omega$ also increase. Increasing $j$ also affects the real parts of $\omega$ such that these become larger. Moreover, as the number of spacetime dimensions increase, Re$(\omega)$ and Im$(\omega)$ increase in magnitude. Therefore, the scalar perturbation modes decay faster with higher $d$ and smaller $a$ values. 

\setlength{\tabcolsep}{15pt}
\begin{table}[h]

\centering

\begin{tabular}{|c|c|c|c|c|}
\hline
$a$ & $j=0$  & $j=1$ & $m\Omega_c$ & $m\Omega_h$  \\  \hline

0 & 1.1463 $-$ 0.4027$i$  & 1.5935 $-$ 0.3971$i$ & 0 & 0  \\
 
\hline

0.1 & 1.1721 $-$ 0.3952$i$   & 1.6082 $-$ 0.3905$i$ & 6.3$\times 10^{-9}$ & 0.0627  \\
 
\hline
 
0.2 & 1.1987 $-$ 0.3905$i$ & 1.6252 $-$ 0.3838$i$ & 1.3$\times 10^{-8}$ & 0.1246   \\
 
\hline

0.3 & 1.2251 $-$ 0.3874$i$ & 1.6441 $-$ 0.3774$i$ & 1.9$\times 10^{-8}$ & 0.1849  \\
 
\hline

\end{tabular}
\caption{The QNMs for scalar perturbation of six dimensional rotating black hole in dS spacetime computed via AIM. With $M=1,\Lambda=0.001,k=0,m=1,\mu=0.1$.}
\label{tab:tab7}
\end{table}

\setlength{\tabcolsep}{15pt}
\begin{table}[h]

\centering

\begin{tabular}{|c|c|c|c|c|}
\hline
$a$ & $j=0$  & $j=1$ & $m\Omega_c$ & $m\Omega_h$\\  \hline

0 & 1.3642 $-$ 0.5144$i$  & 1.8446 $-$ 0.4955$i$ & 0 & 0 \\
 
\hline

0.05 & 1.3782 $-$ 0.5092$i$  & 1.8522 $-$ 0.4914$i$ & 1.3$\times 10^{-5}$ & 0.0315  \\
 
\hline

0.10 & 1.3925 $-$ 0.5049$i$ & 1.8606 $-$ 0.4874$i$ & 2.5$\times 10^{-5}$ & 0.0629   \\
 
\hline

0.15 & 1.4070 $-$ 0.5015$i$ & 1.8697 $-$ 0.4835$i$ & 3.8$\times 10^{-5}$ & 0.0940   \\
 
\hline

0.20 & 1.4215 $-$ 0.4986$i$ & 1.8794 $-$ 0.4797$i$ & 5.1$\times 10^{-5}$ & 0.1246  \\
 
\hline

0.25 & 1.4359 $-$ 0.4964$i$ & 1.8895 $-$ 0.4762$i$ & 6.3$\times 10^{-5}$  & 0.1547  \\
 
\hline

\end{tabular}
\caption{The QNMs for scalar perturbation of seven dimensional rotating black hole in dS spacetime computed via AIM. With $M=1,\Lambda=0.05,k=0,m=1,\mu=0.01$.}
\label{tab:tab8}
\end{table}

\begin{figure}[h]
        \centering
        \includegraphics[width=0.4\textwidth]{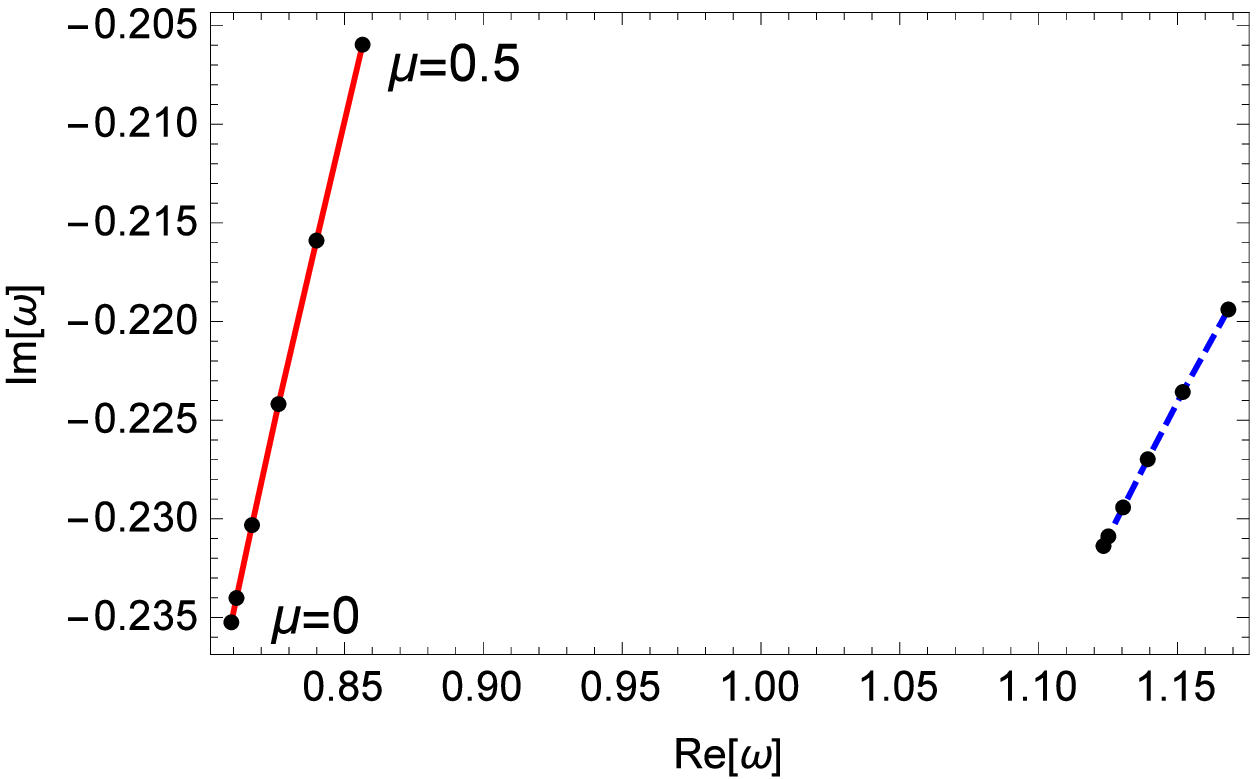}  
        \includegraphics[width=0.4\textwidth]{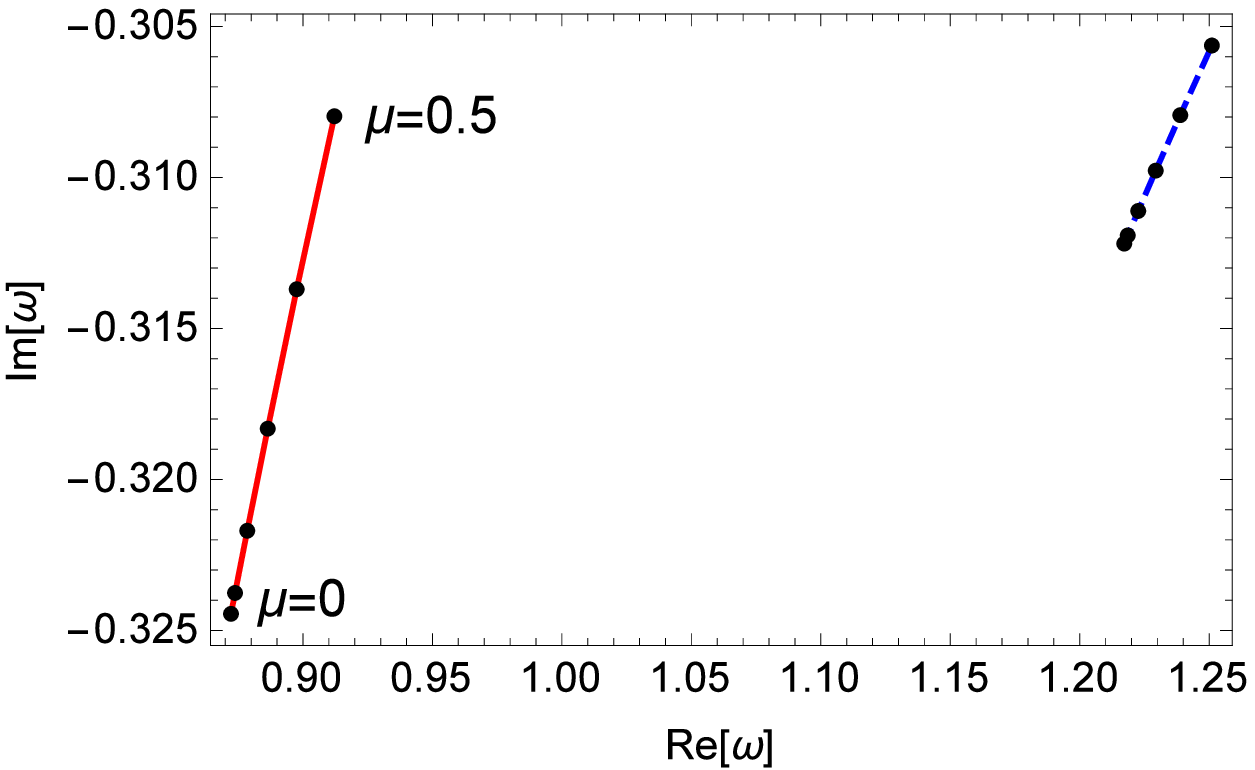}  
        \includegraphics[width=0.4\textwidth]{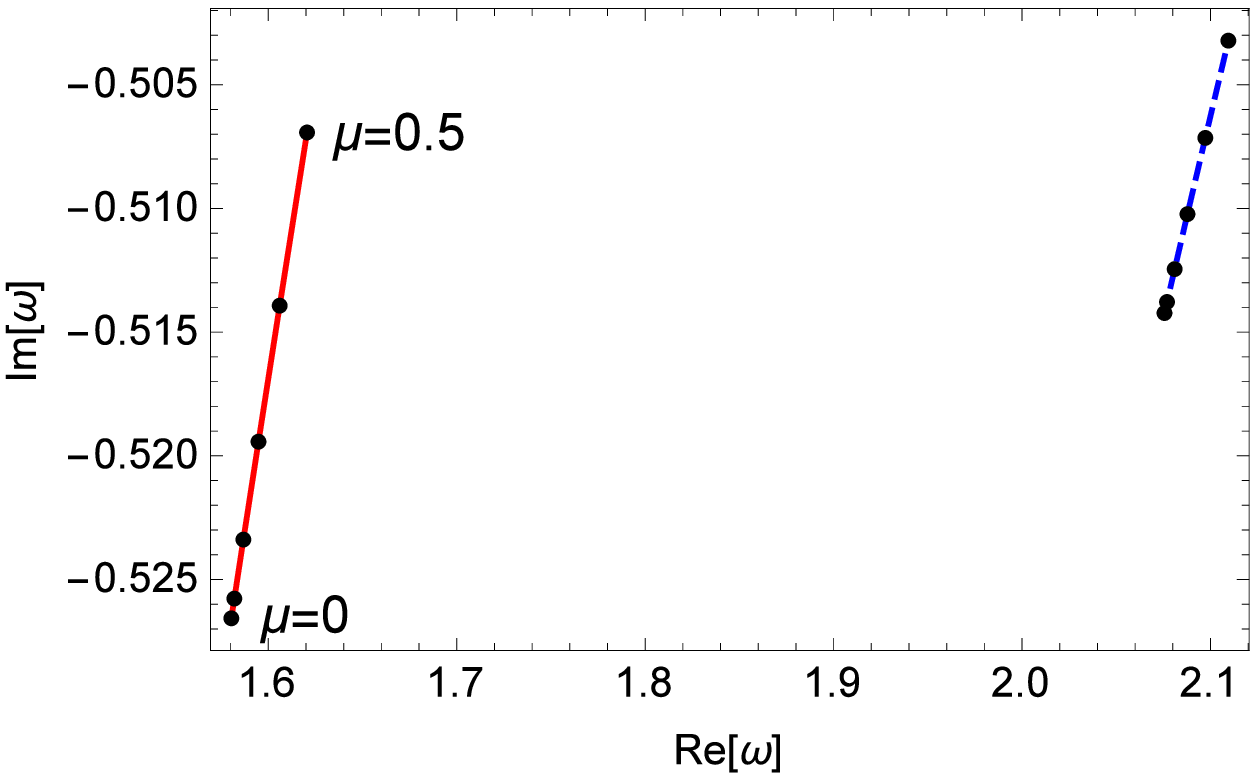}  
        \caption{The QNMs of scalar pertubation of rotating black hole in dS spacetime. The red solid lines are $j=0$ and blue dashed lines are $j=1$ modes. The left upper plot: $d=5,a=0.4,\Lambda=0.001$. The right upper plot: $d=6,a=0.2,\Lambda=0.08$. The bottom plot: $d=7,a=0.15,\Lambda=0.01$.} \label{AIMplotmu}
\end{figure}

In Fig~\ref{AIMplotmu}, the quasinormal frequencies in five, six, and seven dimensions are shown. In these plots, the mass of the black hole is fixed at unity, and the scalar mass $\mu$ varies from $0-0.5$. In each plot, the results are compared between the modes $j=0$ (solid red lines) and $j=1$ (dashed blue lines). The real and imaginary parts of quasinormal frequencies increase as $\mu$ increases. Additionally, the scalar perturbation modes decay slower in massive cases. Note that all the results considered herein are not superradiant modes. Furthermore, all the perturbation modes are exponentially decay.

\subsection{Arbitrary Spin Parameter}

\begin{figure}[h]
\includegraphics[width=0.48\textwidth]{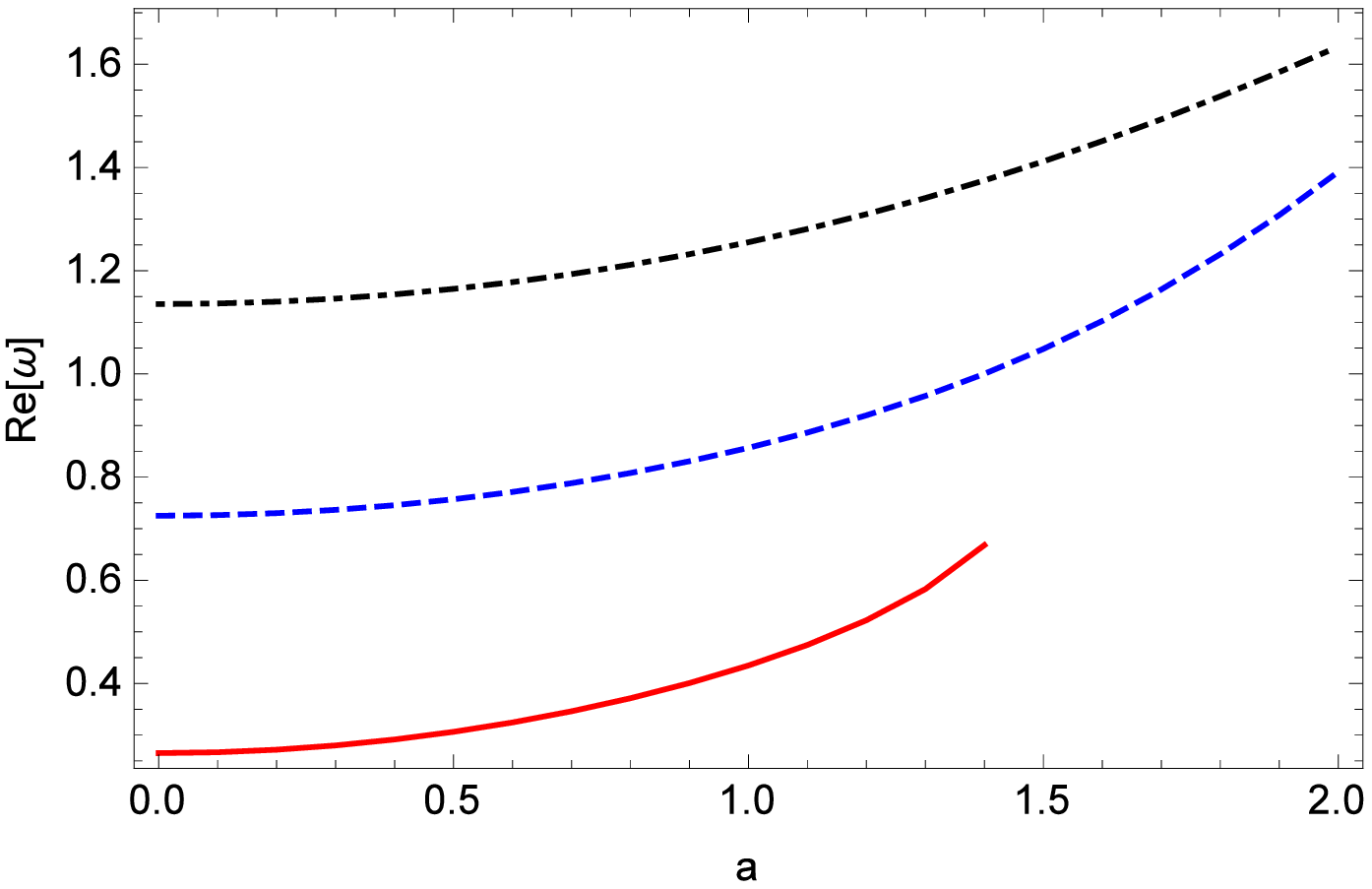}
\includegraphics[width=0.48\textwidth]{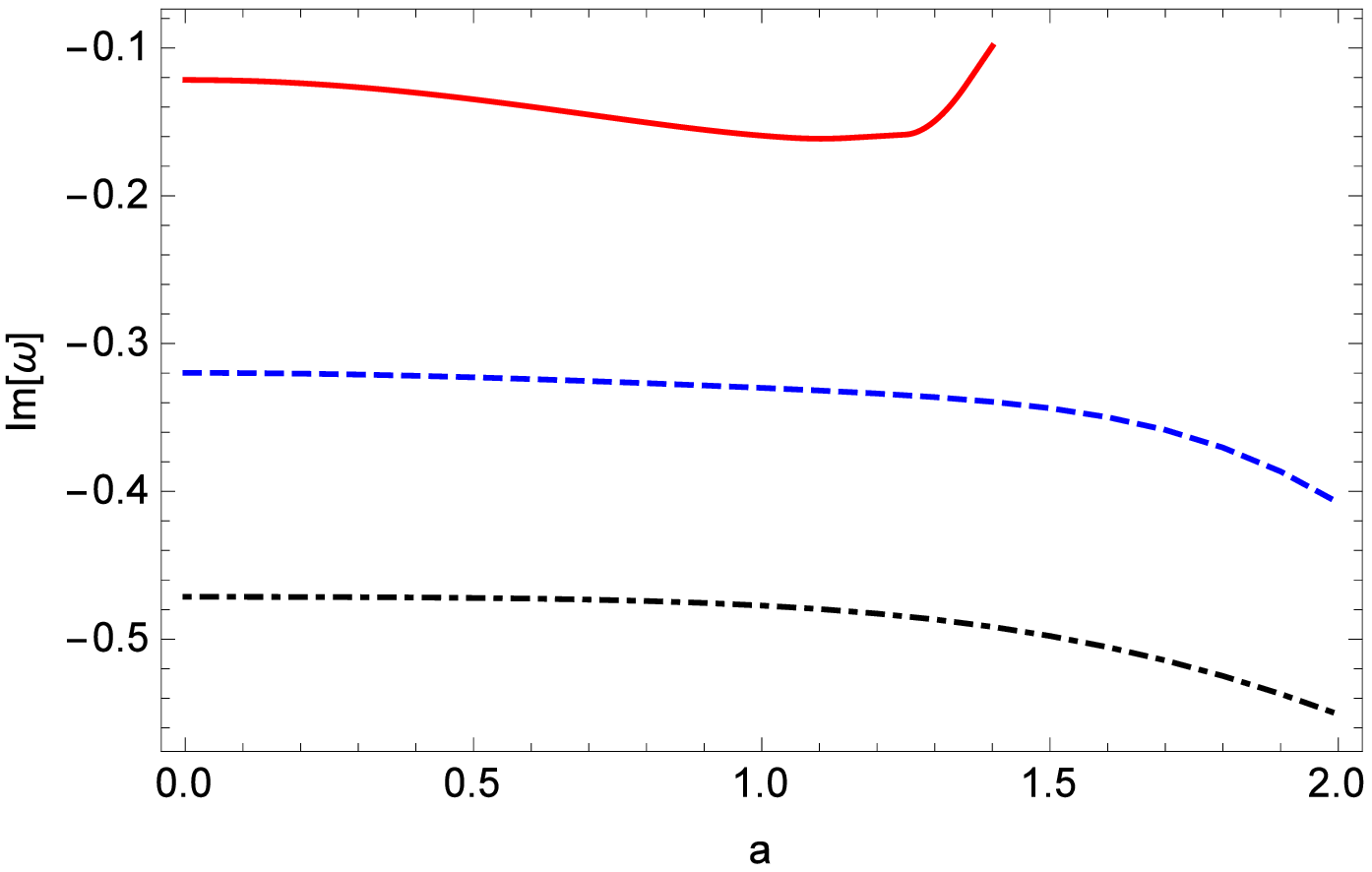}
\caption{The quasinormal frequencies of massless scalar field computed via the AIM for MP-dS black holes. The background parameters are chosen as $M=1,a=0.1,\Lambda=0.1,k=0,j=1,m=0$. (Red, solid): $d=5$. (Blue, dashed): $d=6$. (Black, dotdashed): $d=7$ } 
\label{Extendm0}
\end{figure}

The extension of Table~\ref{tab:tab5} to arbitrary spin parameters is shown in Fig~\ref{Extendm0}. The real and imaginary parts of quasinormal frequencies in $d=5-7$ are plotted against spin parameters. As the spin parameter becomes larger, the energy of the scalar field increases, and Im$(\omega)$ decreases. In addition, for $d=5$, the imaginary part begins to increase after reaching a certain value (in this case, $a=1.2$). Note that this phenomenon is only observed in the five-dimensional case. Moreover, the results for $d=5$ cannot be further extended to a higher spin, because for each fixed $\Lambda$, there is a maximum value of $a$ for which the black hole exists (see Fig~\ref{phasespace}). For a fixed $\Lambda=0.1$, the black hole exists at the largest possible value $a=1.4$. The results for the small $a$ limit are reproduced by the proposed AIM procedure; the findings of the two approaches are observed to be in good agreement.

\begin{figure}[h]

\includegraphics[width=0.48\textwidth]{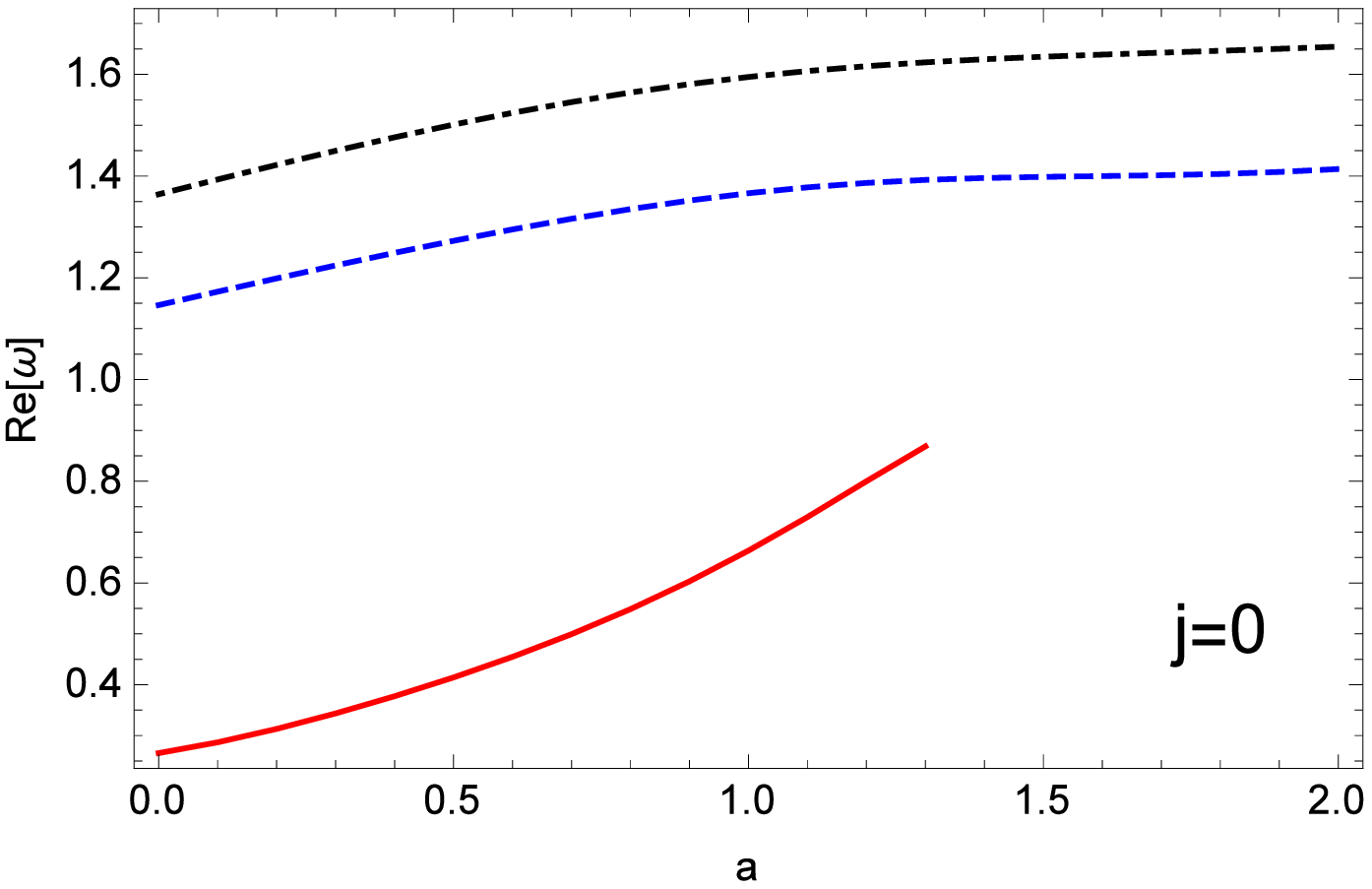}
\includegraphics[width=0.48\textwidth]{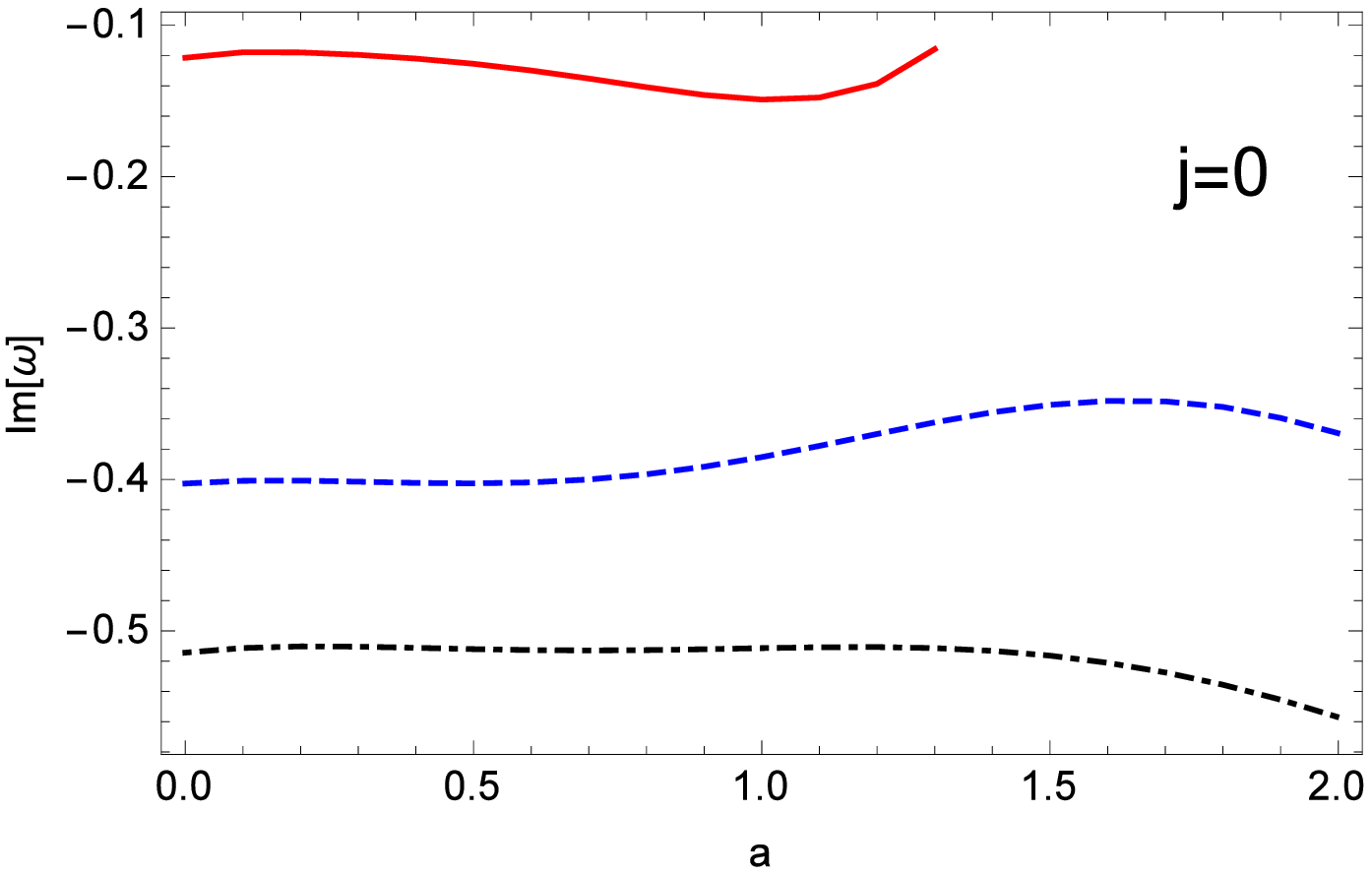}
\includegraphics[width=0.48\textwidth]{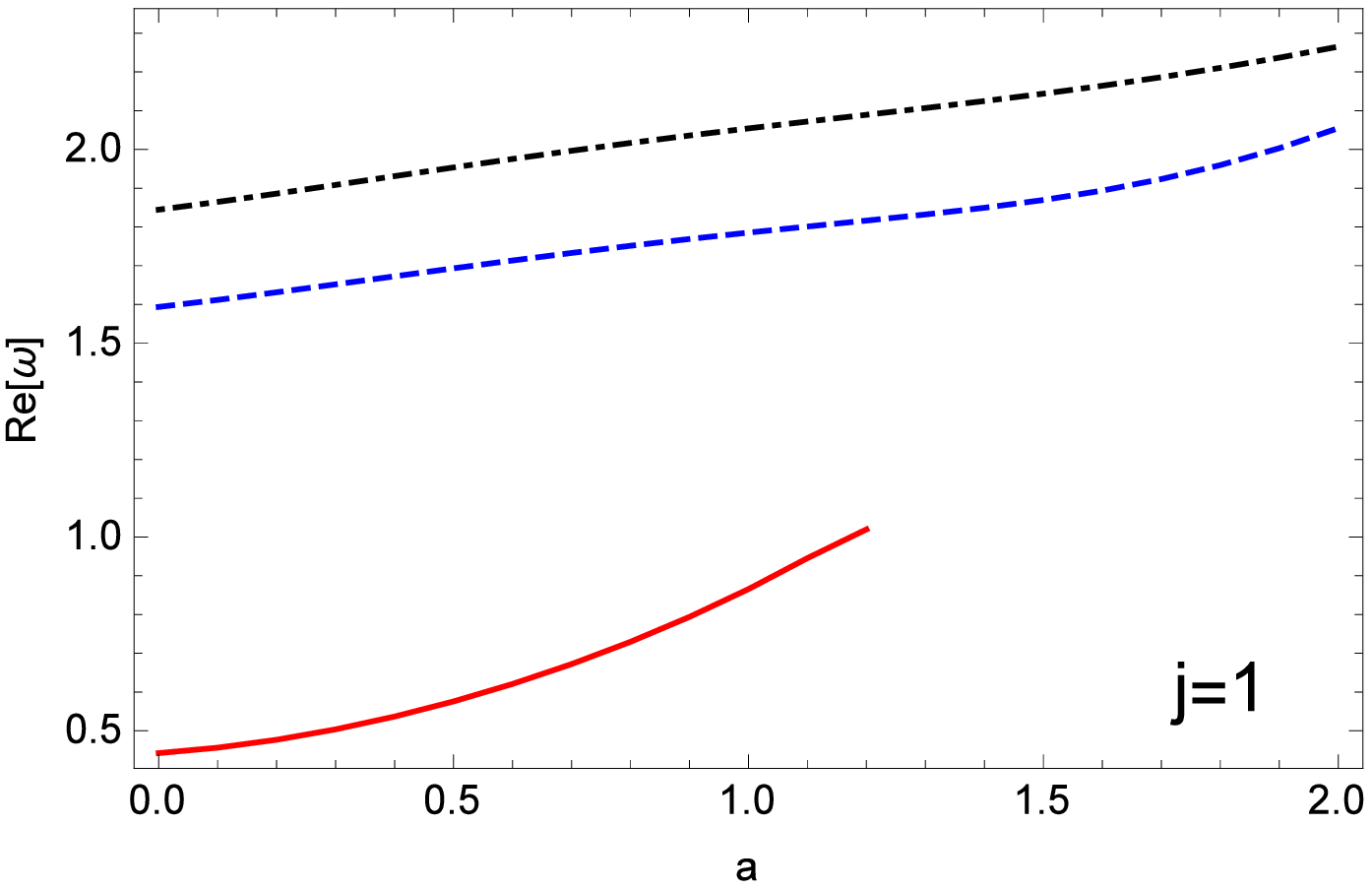}
\includegraphics[width=0.48\textwidth]{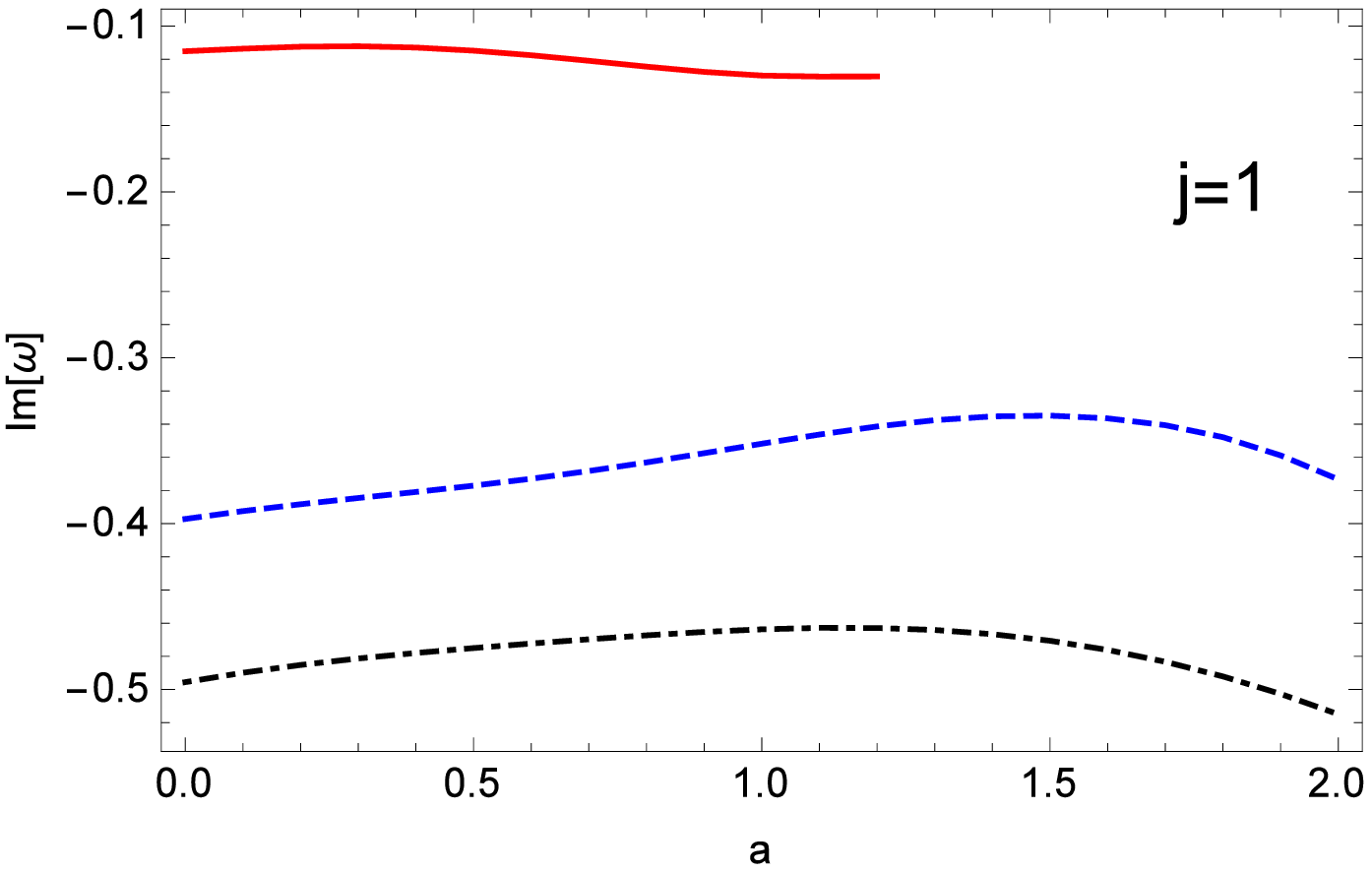}
\caption{The quasinormal frequencies computed via the AIM for MP-dS black holes. The background parameters are chosen as $M=1,k=0,m=1$. (Red, solid): $d=5,\Lambda=0.1,\mu=0.05$. (Blue, dashed): $d=6,\Lambda=0.001,\mu=0.1$. (Black, dotdashed): $d=7,\Lambda=0.05,\mu=0.01$.} 
\label{Extendm1}
\end{figure}

Figure~\ref{Extendm1} shows the comparisons between the $j=0$ and $j=1$ modes of the quasinormal frequencies of massive scalar fields on MP-dS black holes in five, six, and seven dimensions. A similar trend can be observed for Re$(\omega)$. The energy of the scalar perturbation mode increases with $a$ and the eigenvalue of hyperspherical harmonics $j$. Moreover, marginally increasing $j$ increases Im$(\omega)$. For a small $a$ value, the imaginary part of the quasinormal frequency increases with the black holes' spin for $d=5-7$. For a large value of $a$, the black holes become more stable as $a$ increases. However, in five dimensions, Im$(\omega)$ decreases again once it passes a certain point. It is found that the convergence of the proposed method worsens as the ``edge'' of the black hole phase space is approached. Thus, for $d=5$, the results are shown up to $a=1.3$ and $1.2$ for $j=0$ and $j=1$, respectively. Beyond these points, the proposed numerical method fails to produce a converged quasinormal frequency value.

\begin{figure}[h]

\includegraphics[width=0.48\textwidth]{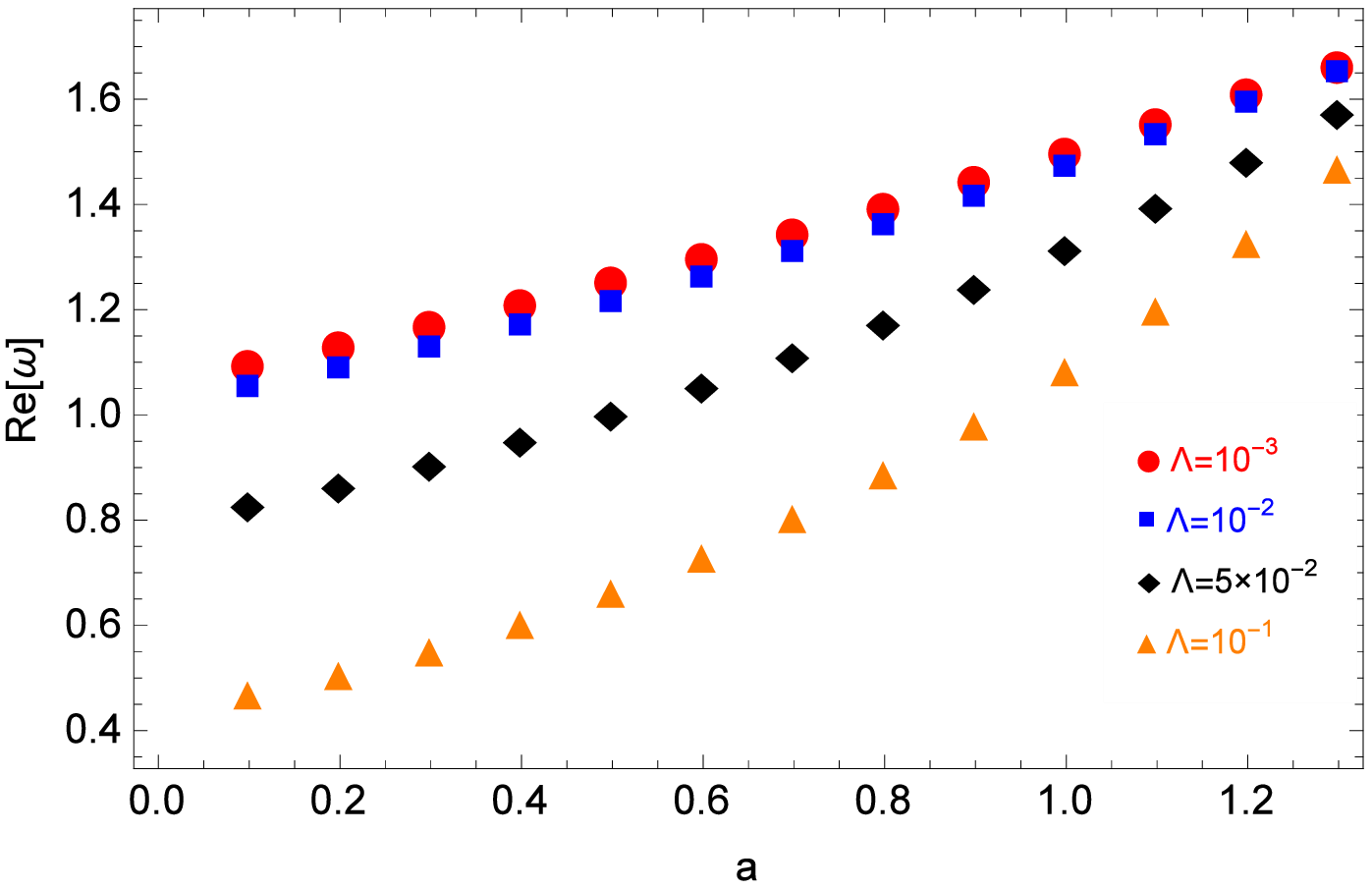}
\includegraphics[width=0.48\textwidth]{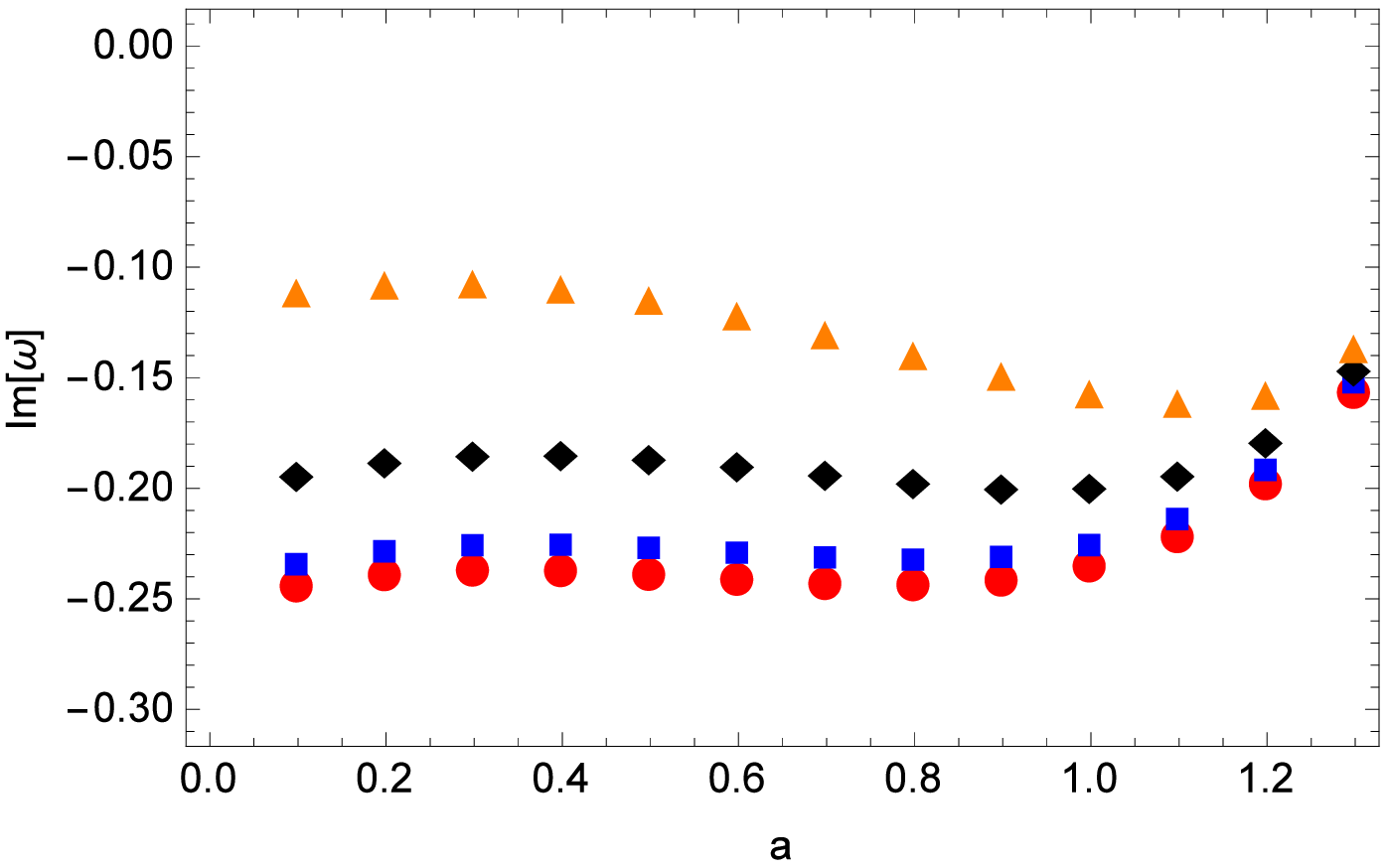}
\caption{The quasinormal frequencies computed via the AIM for five dimensional MP-dS black holes. The background parameters are chosen as $M=1,k=0,j=0,m=2,\mu=0.1$.} 
\label{Extendm2}
\end{figure}

Figure~\ref{Extendm2} shows the effect of the cosmological constant on the quasinormal frequency. The Re$(\omega)$ and Im$(\omega)$ are plotted against the spin parameter of the black hole for various $\Lambda$ values. It is clear that the real parts of $\omega$ decrease with $\Lambda$. In contrast, as $\Lambda$ increases, the imaginary parts also increase. At a significantly small cosmological constant, i.e., $\Lambda=10^{-3},10^{-2}$, increasing $\Lambda$ marginally affects Re($\omega$) and Im$(\omega)$. In contrast, for larger $\Lambda$ values, the quasinormal frequencies vary more significantly as $\Lambda$ increases. Finally, it is noted that the difference of the real (imaginary) parts between each fixed $\Lambda$ is larger for small $a$ values; the difference becomes less significant for a large black hole's spin. 

All the results explored in this section do not satisfy the superradiant condition. Throughout the numerical investigations of this study, no unstable modes were found. All the black holes considered here are found to be stable against linear scalar perturbations.

\section{Summary}\label{sec:conclu}

This study investigated the quasinormal modes of the massive scalar field on MP-dS black holes with a single rotation. At $d=5$, the black holes possess three positive roots, thus having two extremal limits, i.e., $r_C\to r_h$ and $r_h \to r_c$. For a fixed value $M$, there exists a maximum black hole spin parameter value for which black hole solutions exist. In $d\geq 6$, the black hole only has two horizons ($r_h$ and $r_c$), and thus, it has only one extremal scenario. Moreover, singly rotating black holes with dimensions greater than five could acquire arbitrarily large spins. The scalar field equation on curved spacetime can be separated into radial and angular parts in a higher-dimensional singly rotating background. Herein, the results in \cite{Cho:2009wf} were generalized, and an analytic formula of the angular eigenvalue was obtained for a slowly rotating limit. 

This study first considered the metric (\ref{metric}) in the near-extremal limit. In this case, the near-extremal scenario occurs when the event horizon of the black hole is taken to be significantly close to the cosmological constant. This can be achieved by a careful selection of background parameters. Within this limit, the radial equation can be transformed into the wave equation with the P\"oschl-Teller effective potential. When subject to the quasinormal boundary condition, the quasinormal frequencies of a near-extremal black hole can be analytically obtained. The real part of the quasinormal frequency depends on the matter field contents, whereas its imaginary part is proportional to the surface gravity of the black hole. The real and imaginary parts of the near-extremal frequency were found to increase and decrease, respectively, with the number of spacetime dimensions. 

By following the WKB method for Kerr black holes \cite{Seidel:1989bp}, the QNMs of slowly rotating MP-dS with a single rotation parameter were calculated. Seidal's approach was improved by considering the higher order corrections (upto sixth order) \cite{Konoplya:2003ii}. Then, the results from the WKB and P\"oshcl-Teller method within the near-extremal limit were compared. The two methods were found to be in excellent agreement. However, it was observed that the higher correction terms of WKB i.e., $\Lambda_{4-5}$ diverged as the difference between the cosmic and event horizon decreased. Thus, only the third-order WKB is applicable when considering the near-extremal regime of spacetime.

The QNMs of higher-dimensional singly rotating black holes were fully investigated by the improved AIM. The quasinormal frequency $\omega$ and the angular eigenvalue $A_{kjm}$ were numerically solved. The AIM coefficients were analytically derived from the radial and angular equation. The results using the improved AIM, WKB, and P\"oschl-Teller formula were compared in the near-extremal scenarios within a small rotation limit. The three approaches were found to be in good agreement. Beyond the near-extremal limit, the QNMs computed via improved AIM and sixth-order WKB were calculated and shown to be in good agreement. Furthermore, a trend similar to that for a near-extremal limit was observed. As $d$ increased, Re$(\omega)$ increased, whereas Im$(\omega)$ decreased. Note that the deviation of the AIM from the sixth-order WKB became increasingly significant as the dimension of spacetime increased. Additionally, as the size of the scalar field increased, the quasinormal frequency oscillated faster and decayed slower. In a large $a$ regime, it was seen that Re$(\omega)$ increases with $a$ but decreases with $\Lambda$. Unlike the MP-AdS black holes, which are superradiantly unstable, no evidence of any instabilities was found in either small or large $a$. All the perturbation modes explored in this work showed exponential decay. Therefore, these black holes are stable under linear scalar perturbations. Additionally, all the results presented herein do not fulfill the superradiant condition. Thus, they are not superradiant modes.

To fully understand the stability of singly rotating MP-dS black holes, other types of perturbations in this area, i.e., fermionic field, vector, and tensor perturbations, could be considered. Moreover, the stability of multidimensional rotating dS black holes with equal rotation parameters could also be examined by implementing techniques described in this work.

\acknowledgments
This work was supported by the National Research Foundation of Korea (NRF) grant funded by the Korea government (MSIT) (NRF-2018R1C1B6004349) and the Dongguk University Research Fund of 2020.

\end{document}